\documentclass[reprint,pre,aps,twocolumn,showpacs,preprintnumbers,superscriptaddress,a4paper]{revtex4}

\usepackage{graphicx}
\usepackage{dcolumn}
\usepackage{bm}
\usepackage{amssymb}
\usepackage{amsmath}
\usepackage[nooneline]{subfigure}
\usepackage{hyperref}

\usepackage{graphicx,psfrag}
\usepackage{xfrac}

\def\urlprefix{}

\begin{document}

\title{First-order phase transitions in the Kuramoto model\\with compact bimodal frequency distributions}

\author{Bastian Pietras} 
\email{b.pietras@vu.nl}
\affiliation{%
Amsterdam Movement Science \& Institute for Brain and Behavior Amsterdam, Faculty of Behavioural and Movement Sciences, Vrije Universiteit Amsterdam, Amsterdam 1081 BT, The Netherlands}
\affiliation{Department of Physics, Lancaster University, Lancaster LA1 4YB, United Kingdom}

\author{Nicol\'as Deschle}%
\affiliation{%
 Amsterdam Movement Science \& Institute for Brain and Behavior Amsterdam, Faculty of Behavioural and Movement Sciences, Vrije Universiteit Amsterdam, Amsterdam 1081 BT, The Netherlands}
\affiliation{Institute for Complex Systems and Mathematical Biology, University of Aberdeen, King's College, Old Aberdeen AB24 3UE, United Kingdom}

\author{Andreas Daffertshofer}
\email{a.daffertshofer@vu.nl}
\affiliation{%
 Amsterdam Movement Science \& Institute for Brain and Behavior Amsterdam, Faculty of Behavioural and Movement Sciences, Vrije Universiteit Amsterdam, Amsterdam 1081 BT, The Netherlands}

\date{\today}

\begin{abstract}
The Kuramoto model of a network of coupled phase oscillators exhibits a first-order phase transition when the distribution of natural frequencies has a finite flat region at its maximum.
First-order phase transitions including hysteresis and bistability are also present if the frequency distribution of a single network is bimodal.
In this study we are interested in the interplay of these two configurations and analyze the Kuramoto model with compact bimodal frequency distributions in the continuum limit.
As of yet, a rigorous analytic treatment has been elusive.
By combining Kuramoto's self-consistency approach, Crawford's symmetry considerations, and exploiting the Ott-Antonsen ansatz applied to a family of rational distribution functions that converge towards the compact distribution, we derive a full bifurcation diagram for the system's order parameter dynamics.
We show that the route to synchronization always passes through a standing wave regime when the bimodal distribution is compounded by two unimodal distributions with compact support.
This is in contrast to a possible transition across a region of bistability when the two compounding unimodal distributions have infinite support.

\end{abstract}
\maketitle

\section{Introduction}
Spontaneous synchronization is an omnipresent phenomenon in nature.
Be it the flashing of fireflies, epileptic seizures in the brain, or cascading failures in power grids -- real word systems can exhibit abrupt transitions from incoherence to synchrony
\cite{Strogatz2003Sync,boccaletti2006complex}.
A mathematical approach to understand the mechanisms behind such first-order phase transitions, also known as explosive synchronization
\cite{arenas2008synchronization,Zhang2015explosive}, commonly dwells on the paradigmatic and analytically tractable Kuramoto model of coupled phase oscillators \cite{Kuramoto1984,Strogatz2000,Acebron_Review2005}.
First-order phase transitions in the Kuramoto model have first been reported by Paz\'o in \cite{Pazo2005}.
He considered a uniform distribution of the oscillators' natural frequencies, which led to a discontinuous phase transition from incoherence to synchronization in a network of infinitely many oscillators.
Basnarkov and Urumov extended Paz\'o's results to all frequency distributions that have a plateau at their maximum \cite{basnarkov2007}.
They argued that their frequency distributions represent an intermediate case between unimodal and bimodal distributions.

The Kuramoto model with bimodal frequency distributions is long known to exhibit first-order phase transitions including hysteresis and bistability \cite{bonilla1992nonlinear}.
Symmetric bimodal frequency distributions already allow for a wider range of bifurcations off the incoherent solution, giving rise to both steady-state and oscillatory solution branches.
The bimodal Kuramoto model has therefore been in the focus of investigation for decades \cite{Kuramoto1984,bonilla1992nonlinear, Crawford1994,bonillaperezvicente1998,Montbrio2006}.
The introduction of the seminal Ott-Antonsen (OA) ansatz \cite{OttAntonsen2008,OttAntonsen2009} paved the way to rigorously derive the full spectrum of collective dynamics of the bimodal Kuramoto model \cite{MartensExactResults2009,PazoMontbrio2009existence}.
Remarkably, first-order phase transitions naturally appear in the system with bimodal distributions in the absence of a flat region.

Despite the progress in describing the collective dynamics of the Kuramoto model, there remain unsolved problems.
One of these is given by the combination of bimodality and compact support of the frequency distribution.
The non-analyticity of the uniform frequency distribution with compact support, hereafter referred to as a compact distribution, no longer renders the OA ansatz applicable.
To overcome this difficulty, a direct approach is to rely on Kuramoto's original self-consistency argument \cite{Kuramoto1984}.
By determining so the parameter regions of either incoherent or (partially) synchronous solutions, this approach reveals the backbone of the bifurcation diagram.
Yet, it also raises new issues -- let alone that the offset of the well-reported oscillatory solution branches off the incoherent solution cannot be determined.
An alternative strategy is to dwell on the recently introduced family of rational distribution functions that converge to the compact distribution and that are applicable to the OA ansatz \cite{Skardal2018}.
Eventually, referring to Crawford's intuition about the role of symmetries in the system \cite{Crawford1994} may fill the missing gaps.

Combining the different analytic approaches, we here present a full bifurcation diagram of the Kuramoto model with a bimodal frequency distribution that is compounded by two compact distributions.
Moreover, we answer whether the nature of phase transitions changes when the bimodal frequency distributions exhibits two (symmetric) plateaus.
It is instructive to contemplate the dynamical regimes of the bimodal Kuramoto model and the various routes of synchronization. 
In general, the possible regimes are incoherence (``no sync''), partial synchronization (``partial sync''), and macroscopic oscillations -- Crawford coined this solution a standing wave \cite{Crawford1994}.
In fact, there may exist regions of bistability between a partially synchronized state and either the incoherent solution or macroscopic oscillations.
When increasing the coupling strength subsequently, the transitions between distinct collective behavior are the following.

\vspace*{1em}\noindent
(a) For a bimodal distribution as the \emph{sum of two} even unimodal distributions \cite{MartensExactResults2009}:
\begin{enumerate}
		\item[(a1)] \emph{``No sync $\to$ oscillations $\to$ partial sync''},\\
		if the two peaks of the distributions are well separated.
		\item[(a2)] \emph{``No sync $\to$ bistability $\to$ partial sync''},\\
		if the two peaks of the distributions are sufficiently near.
\end{enumerate}
(b) For a bimodal distribution as the \emph{difference of two} even unimodal distributions \cite{PazoMontbrio2009existence}:
\begin{enumerate}
		\item[(b1)] \emph{``No sync $\to$ bistability $\to$ partial sync''},\\
		if the central dip between the two peaks of the distributions does not reach zero.
		\item [(b2)] \emph{``No sync $\to$ oscillations $\to$ partial sync''},\\
		if the central dip between the two peaks reaches zero.
\end{enumerate}
In case (b) the two peaks of the distribution can become arbitrarily close to each other for finite widths of the respective unimodal distributions.
That is, while in case (a) there is always a parameter region of bistability close to the unimodal-bimodal border, scenario (b2) shows that this proximity does not necessarily imply hysteresis \cite{PazoMontbrio2009existence}.
In our study, we show that also in case (a) of the sum of two unimodal but compact distributions the route of synchronization via a bistable regime no longer exists.
Moreover, the phase transition from incoherence to collective oscillations is of first-order, which underlines the well-known nature of explosive synchronization for compact frequency distributions \cite{Pazo2005}.
However, the compact bimodal distribution can equally be written as the sum but also as the difference of two unimodal compact distributions.
It thus defines a natural limit of both sum and difference formulations, and can be considered a descriptive link from one to another route of synchronization.

The paper is organized as follows.
In \textit{Section}~\ref{sec:KM} we briefly review the Kuramoto model and the different analytic approaches to analyze the collective behavior.
We exemplarily apply the techniques to the unimodal Kuramoto model with rational frequency distributions in \textit{Section}~\ref{sec:unimodal} and present the results in an intuitive way.
Subsequently, we consider bimodal rational frequency distributions in \textit{Section}~\ref{sec:bimodal_rational} and provide complete bifurcation diagrams for these flat bimodal Kuramoto models.
In \textit{Section}~\ref{sec:bimodal_compact} we turn to the compact bimodal Kuramoto model and combine different analytic and numeric approaches to describe the full dynamical spectrum.
Finally, we investigate the different routes to synchronization in \textit{Section}~\ref{sec:explosiveOrcontinuous}.
In \textit{Section}~\ref{sec:conclusion} we summarize and discuss our results.

\section{Collective behavior of the Kuramoto model}
\label{sec:KM}

We consider globally coupled phase oscillators and trace the macroscopic behavior in terms of the order parameter dynamics.
The dynamics of each oscillator $k$ is given by
\begin{equation}\label{eq:bimodal_microdynamics}
\dot{\theta}_k = \omega_k + \frac{K}{N} \sum_{l=1}^N \sin(\theta_l - \theta_k) \ , \quad k = 1, \dots , N,
\end{equation}
where the natural frequency term $\omega_k$ is drawn from a distribution density $g(\omega)$.
With the Kuramoto order parameter
\begin{equation}\label{eq:Kuramoto_OP}
z = R \mathrm{e}^{i\Psi} = \frac{1}{N} \sum_{k=1}^N \mathrm{e}^{i\theta_k} \ ,
\end{equation}
one can rewrite \eqref{eq:bimodal_microdynamics} as 
\begin{equation}\label{eq:bimodal_microdynamics1}
\dot{\theta}_k = \omega_k + K R \sin(\Psi - \theta_k) \ .
\end{equation}
The amplitude $R \in [0,1]$ indicates the degree of global synchronization and, together with the coupling strength $K$, scales the diffusive coupling of each oscillator with respect to the mean phase $\Psi \in [0, 2\pi)$.

\subsection{Kuramoto's self-consistency analysis}\label{subsec:self-cons_general}
In \cite{Kuramoto1984} Kuramoto presented a self-consistency argument to determine the critical coupling strength at the transition from one to another collective behavior.
In brief, the solutions of \eqref{eq:bimodal_microdynamics1} typically exhibit two types of long-term behavior.
Those oscillators with frequency $|\omega_k| \leq KR$ approach a stable fixed point and are `phase locked' according to
\begin{equation}\label{eq:phaselocked}
\omega_k = KR \sin(\theta_k) ;
\end{equation} 
note that one can set $\Psi=0$ due to the rotational symmetry of the system.
On the other hand, oscillators with $|\omega_k| > KR$ are `drifting' and progress around the circle.
In view of the seeming contradiction that the order parameter can be constant in spite of a drifting fraction of oscillators, Kuramoto used \eqref{eq:phaselocked} to reduce the self-consistency equation
\begin{equation}\label{eq:self-cons}
R= KR \int_{-\pi/2}^{\pi/2} \cos^2(\theta) g(KR \sin \theta) \ d\theta \ .
\end{equation}
\eqref{eq:self-cons} always has the trivial solution $R=0$ for any coupling strength $K$.
However, there is a second non-trivial solution branch that satisfies
\begin{equation}\label{eq:self-consK}
1= K \int_{-\pi/2}^{\pi/2} \cos^2(\theta) g(KR \sin \theta) \ d\theta \ .
\end{equation}
This solution bifurcates off $R=0$ at a critical value $K = K^\ast = 2/\big[\pi g(\Omega)\big]$, where $g(\omega)$ is symmetric about the central frequency $\Omega$.

For certain frequency distributions $g(\omega)$, the integral in \eqref{eq:self-consK} can be evaluated explicitly.
For other frequency distributions, analytic expressions become more involved and one may have to rely on numerics.
In any case, the self-consistency equation \eqref{eq:self-cons} determines the asymptotic value of a fixed point solution of the collective dynamics.
However, this approach does not reveal any stability properties about the respective solutions.

\subsection{Strogatz and Mirollo's density approach}
A mathematically sound description of the problem of stability has been depicted first in \cite{StrogatzMirollo1991}.
Considering the continuum limit of infinitely many oscillators, $N \to \infty$, Strogatz and Mirollo introduced a distribution density of oscillators $f(\omega,\theta,t)$ such that for each natural frequency $\omega$, $f(\omega,\theta,t) d\theta$ denotes the fraction of oscillators with this particular natural frequency that lie between phases $\theta$ and $\theta+d\theta$ on the circle at time $t$.
The evolution of $f$ is governed by the continuity equation $\partial_t f + \partial_\theta (\dot \theta f) = 0$, with $\dot\theta$ the continuum version of \eqref{eq:bimodal_microdynamics1}.
The Kuramoto order parameter \eqref{eq:Kuramoto_OP} is now given by 
\begin{equation}\label{eq:orderparameter}
z = R \mathrm{e}^{i\Psi} = \int_{0}^{2\pi} \int_{-\infty}^\infty \mathrm{e}^{i\theta} f(\omega,\theta, t) g(\omega) \ d\omega d\theta
\end{equation}
and the continuity equation becomes \cite{StrogatzMirollo1991}
\begin{equation}\label{eq:strogatzmirollo_continuityeq}
\begin{aligned}
0 &= \partial_t f + \partial_\theta \Big[ f \Big( \omega \ + \\ 
&K \int_{0}^{2\pi} \int_{-\infty}^\infty\sin(\theta'-\theta) f(\omega',\theta', t) g(\omega') \ d\omega d\theta' \Big) \Big].
\end{aligned}
\end{equation}
This nonlinear integro-differential equation for $f$ is the continuum limit of the Kuramoto model \eqref{eq:bimodal_microdynamics}, and it contains all information about the existence, stability and bifurcations of all possible solutions.
The simplest state is the incoherent solution, $f(\omega,\theta,t) \equiv 1/2\pi$, corresponding to the $R=0$ solution, which was found to exhibit a bifurcation at a critical coupling $K^\ast$.
Yet, determining its linear stability properties via \eqref{eq:strogatzmirollo_continuityeq} is already fairly involved \cite{StrogatzMirolloMatthews1992}.

\subsection{Ott-Antonsen ansatz}
A major breakthrough in analyzing the collective dynamics of the Kuramoto model could be achieved in 2008, when Ott and Antonsen published an ingenious idea \cite{OttAntonsen2008}.
They considered the Fourier expansion of $f$ of the form
\begin{equation}
f(\omega,\theta,t) = \frac{1}{2\pi}\left[ 1 + \sum_{k=1}^\infty \hat{f}_k(\omega,t)\mathrm{e}^{in\theta} + \mathrm{c.c.}\right],
\end{equation}
and suggested the ansatz $\hat{f}_k(\omega,t)=\alpha^k(\omega,t)$ for an analytic function $\alpha(\omega,t)$.
In this case, the distribution density $f$ collapses to a Poisson kernel and the corresponding Ott-Antonsen (OA) manifold was proven to define a global attractor of the collective dynamics \cite{OttAntonsen2009}.
On this OA manifold the evolution of the order parameter can be found to exactly follow a certain low-dimensional dynamics.

In more detail, $\alpha$ satisfies \cite{OttAntonsen2009}
\begin{equation}\label{eq:alphaOA}
\partial_t \alpha + i \omega \alpha + \tfrac{K}{2} \big( z \alpha^2 - z^\ast \big) = 0.
\end{equation}
We can close the equation by inserting the OA ansatz into \eqref{eq:orderparameter}, so that the Kuramoto order parameter is given by
\begin{equation}\label{eq:zOA}
z^\ast = \int_{-\infty}^\infty \alpha(\omega,t) g(\omega) d\omega \ .
\end{equation}
Equations \eqref{eq:alphaOA} and \eqref{eq:zOA} define a system of differential equations that exactly describe the order parameter dynamics of the network.
To incorporate \eqref{eq:zOA} into \eqref{eq:alphaOA}, we have to evaluate the integral by using Cauchy's residue theorem.
That is, one has to find the poles $\hat\omega$ of $g(\omega)$ with negative but finite imaginary part and their respective residues $\mathrm{Res}(g;\hat\omega)$.
This step imposes an important analyticity condition on the frequency distribution $g(\omega)$ in that it needs to have a finite number of poles with finite but non-vanishing imaginary part, see also \cite{OttAntonsen2009,OttAntonsenComment} for more details.

When evaluating \eqref{eq:zOA} at the respective poles $\hat\omega_k$, $k=1,\dots,m<\infty$, of $g(\omega)$, we can define corresponding `pseudo' order parameters $z_k$ with $k=1,\dots,m$ by means of
\begin{equation}
z_k^\ast = -2\pi i \mathrm{Res}(g; \hat{\omega}_k) \alpha(\hat{\omega}_k, t) \ ,
\end{equation}
and $z = \sum_{k=1}^m z_k$.
Hence, we find \cite{Skardal2018}
\begin{equation}\label{eq:lowdimdynamics}
\begin{aligned}
\dot{z}_k = &\ i \hat{\omega}_k^\ast z_k + \frac{K}{2} \left[ 2\pi i \big(\mathrm{Res}(g; \hat{\omega}_k)\big)^\ast \Big( \sum_{j=1}^{m} z_j \Big) \right.\\
& \left.- \Big( \sum_{j=1}^{m} z_j^\ast \Big) \frac{z_k^2}{2\pi i \big(\mathrm{Res}(g; \hat{\omega}_k)\big)^\ast} \right],
\end{aligned}
\end{equation}
which exactly describes the collective dynamics of the Kuramoto model for any frequency distribution $g(\omega)$ with a finite number of poles in the lower complex half-plane.

\section{Collective dynamics for flat unimodal frequency distributions}
\label{sec:unimodal}
The unimodal Kuramoto model is well-known to display a phase transition from incoherent to (partially) synchronized collective behavior when the frequency distribution $g(\omega)$ is smooth, even, i.e. symmetric about a central frequency $\Omega$, and decreases for $|\Omega - \omega| > 0$.
The rotational symmetry of the system allows to shift the central frequency to zero and to consider $\Omega=0$.
In this co-rotating frame, the mean phase $\Psi$ of the Kuramoto order parameter, $z=R\mathrm{e}^{i\Psi}$, is constant and one can focus on the amplitude $R=|z|$.
At the critical coupling strength $K^\ast$, a partially synchronized solution with $R>0$ branches off the incoherent solution $R=0$.
As this is a steady-state bifurcation in the co-rotating frame, we can apply Kuramoto's self-consistency approach to determine the non-trivial solution branch.
On the other hand, in the original dynamics, the bifurcation is of Hopf-type and we have to rely on the OA ansatz.

\begin{figure}[t!]
\centering{
\includegraphics[width=\columnwidth]{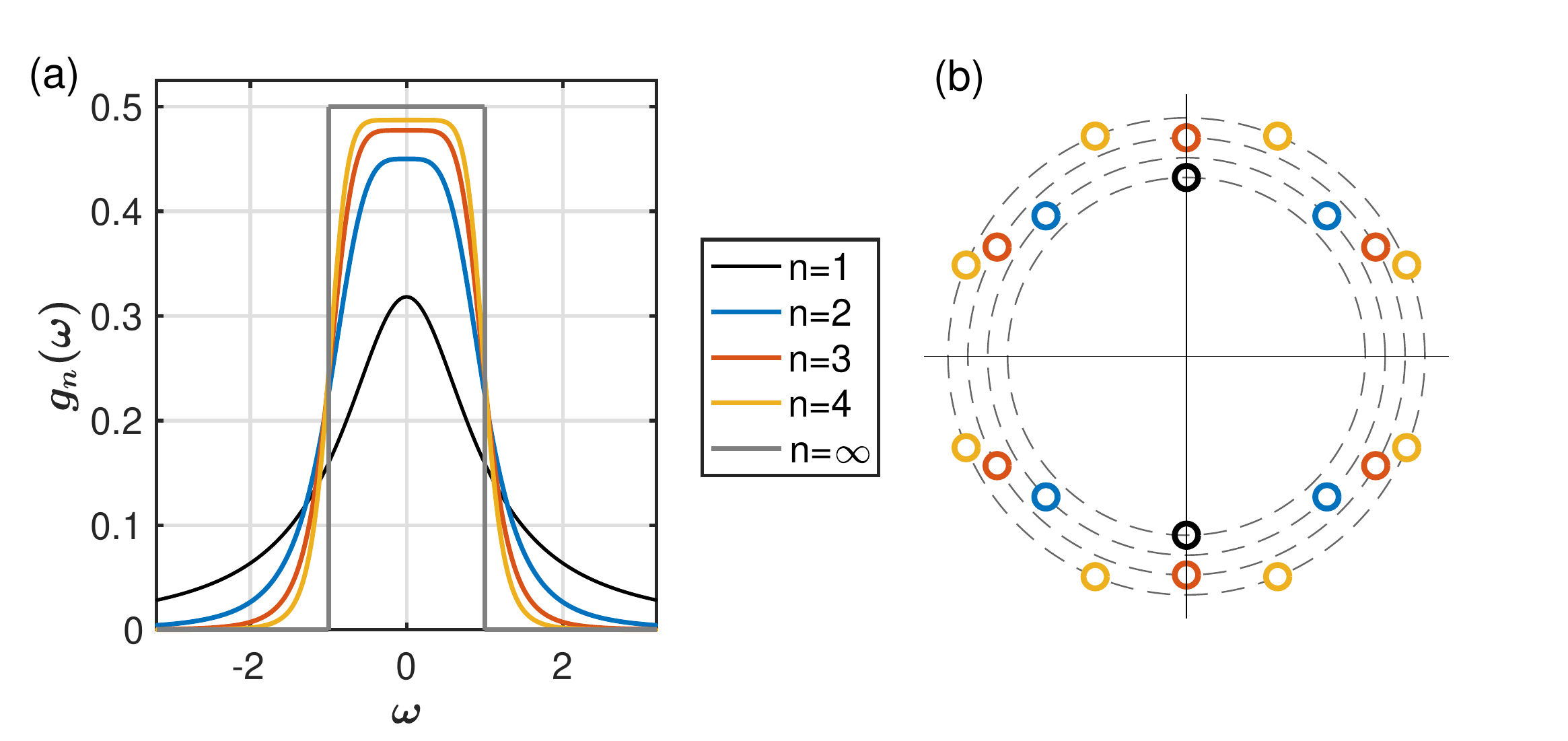}}
\caption{(a) Rational frequency distributions $g^{(1,0)}_n(\omega)$ for $n=1,2,3,4,\infty$ given by \eqref{eq:rationalfrequency}.
(b) Poles of $g^{(1,0)}_n(\omega)$ on the complex unit circle (we scaled the radii for illustration).}
\label{fig:unimodalAll}
\end{figure}

Next, we will briefly revisit the collective dynamics of the unimodal Kuramoto model with flat frequency distributions and compare the different approaches.
We consider the family of rational frequency distributions, $g_n(\omega)$, with $n\in \mathbb{N}$, given by
\begin{equation}\label{eq:rationalfrequency}
g_n(\omega) = g_n^{(\Delta,\Omega)}(\omega)=  \frac{n \sin(\pi/2n)}{\pi} \frac{\Delta^{2n-1}}{(\omega-\Omega)^{2n} + \Delta^{2n}}  \ .
\end{equation}
$\Delta \geq 0$ denotes the half-width at half the height of the distribution, and $n$ defines the order of the polynomial characterizing the frequency distribution.
As $n$ increases, the flat plateau of $g_n(\omega)$ becomes larger and the edges more prominent.
For $n \to \infty$, the rational distribution converges to the compact distribution
$$
\lim_{n\to \infty} g_n(\omega) = g_c(\omega) = \begin{cases} \tfrac{1}{2\Delta} \ , \quad &\text{for } \omega \in [\Omega-\Delta,\Omega + \Delta] \\
0 \ , &\text{otherwise}. \end{cases}
$$
In Fig.~\ref{fig:unimodalAll}(a) we sketch the first four functions of the family of rational distributions $g_n(\omega)$ together with the compact, uniform distribution $g_c$ in the case of $\Delta=1$ and $\Omega=0$.
While the OA ansatz is applicable to rational frequency distributions $g_n(\omega)$ with finite $n<\infty$, the necessary analyticity condition is no longer fulfilled for the compact distribution.

\subsection{Self-consistency approach}\label{subsec:self-cons_unimodal}
\begin{figure}[h!]
\centering{
\includegraphics[width=.85\columnwidth]{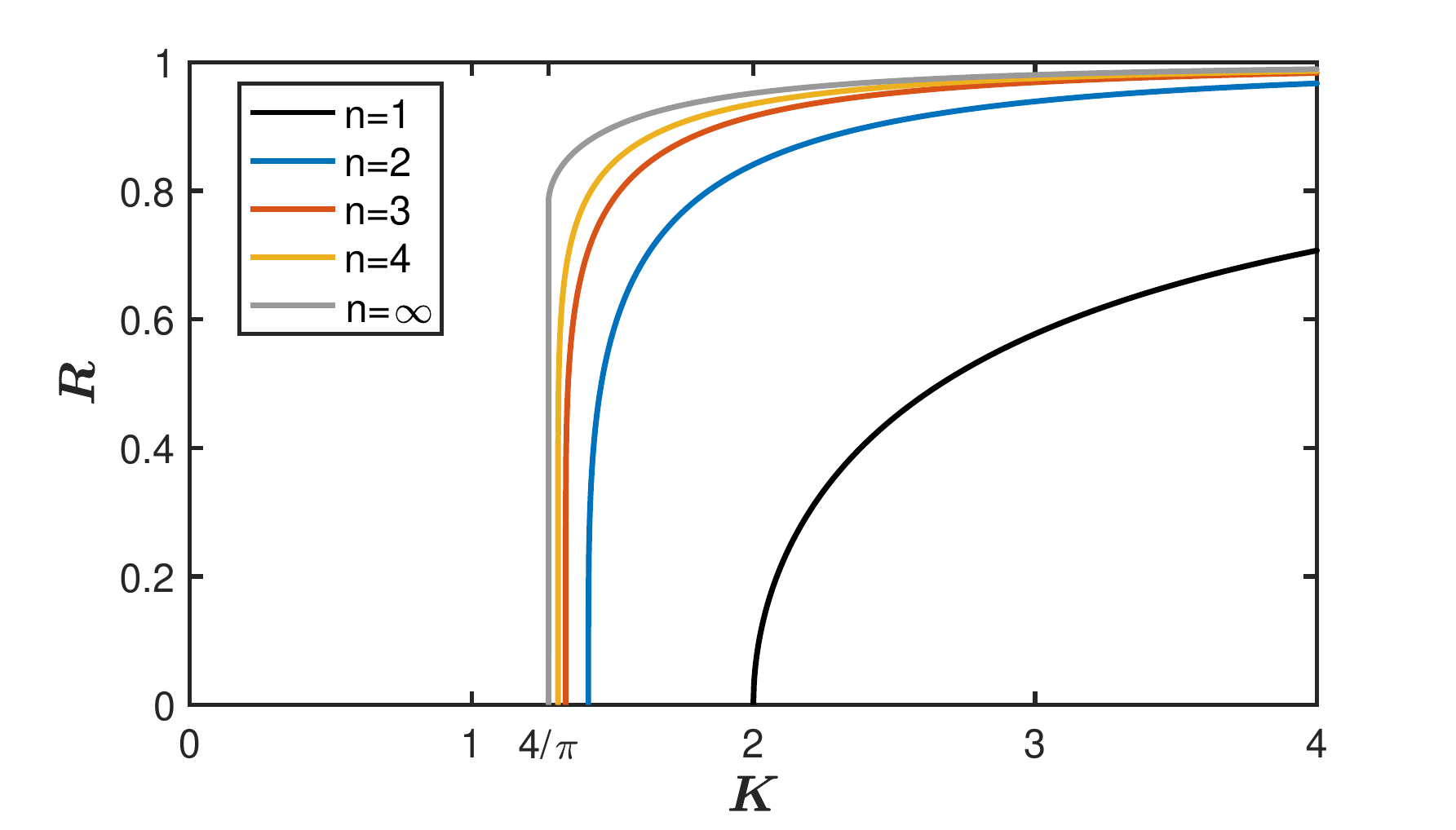}
}
\caption{Towards first-order phase transitions with unimodal rational frequency distributions, $n=1-4$, and the compact distribution $n=\infty$ for $\Delta=1$. The critical coupling strength, at which the nontrivial solution bifurcates off the zero solution, is given by $\kappa_n=2/n\sin(\pi/2n))$.}
\label{fig:ExplosiveSync}
\end{figure}

As mentioned above, the self-consistency approach allows only for certain frequency distributions an explicit solution, while for other distributions the implicit equation \eqref{eq:self-consK} can only be solved numerically.
For $n=1$, the rational frequency distribution is the Lorentzian distribution and \eqref{eq:self-consK} reveals the nontrivial solution $R=\sqrt{1-K^\ast/K}$.
For larger $n>1$, we have to solve \eqref{eq:self-consK} numerically.
The compact distribution, $g_c(\omega)$, however, presents an exception, and \eqref{eq:self-consK} becomes \cite{Pazo2005}
\begin{equation}\label{eq:self-consCompact}
R = \tfrac{1}{2} \sqrt{1 - \big(\tfrac{\Delta}{KR}\big)^2} + \tfrac{KR}{2\Delta} \arcsin\big(\tfrac{\Delta}{KR}\big) \ .
\end{equation}
A solution $R\in \mathbb{R}$ exists only for $KR \geq \Delta$, and $K_cR_c=\Delta$ defines the critical value at which the nontrivial solution branch jumps off the zero-solution from $R=0$ to $R_c = \pi/4$.
For larger coupling $K \geq K_c=4\Delta/\pi$, \eqref{eq:self-consCompact} can be solved implicitly to find a solution $R$, which converges to $1$ for $K\to \infty$.
Fig.~\ref{fig:ExplosiveSync} shows the bifurcation diagrams for the Kuramoto model with unimodal distributions whose plateau size increases successively.
While for $n=1$ the phase transition is of second-order, for larger $n>1$ the phase transitions become more and more discontinuous, culminating in the explosive synchronization behavior of the compact Kuramoto model \cite{Pazo2005}.

\subsection{OA ansatz for rational frequency distributions}
When applying the OA ansatz to the Kuramoto model with rational frequency distributions $g_n(\omega)$, one has to substitute the complex poles of \eqref{eq:rationalfrequency} and their residue values in the reduced order-parameter dynamics (\ref{eq:alphaOA}~\&~\ref{eq:zOA}).
For every $n\in \mathbb{N}$, $g_n(\omega)$ has $n$ simple poles $\hat{\omega}_k$ in the lower complex half-plane.
Then, system (\ref{eq:alphaOA}~\&~\ref{eq:zOA}) becomes $n$-dimensional.
The poles of $g^{(\Delta,\Omega)}_n(\omega)$ are given by $\hat{w}_k = \Omega+\Delta \mathrm{e}^{-i\phi_k}$ with $\phi_k=(2(k-1)+1)\pi/(2n)$, $k=1, \dots, n$, and the respective residues $r_{n,k}$ of the rational distribution function $g_n(\omega)$ at these poles are $r_{n,k}=\mathrm{Res}(g_n; \hat{w}_k)$.
In Fig.~\ref{fig:unimodalAll}(b) we depict the poles of the first four rational distribution functions on the unit circle (the radius is $\Delta=1$) in the complex plane; we have scaled the radii to improve illustration.

In the following, we briefly sketch the analysis by Skardal \cite{Skardal2018} and rephrase his results in an intuitive way highlighting the Hopf character of the bifurcation.
We consider the case $\Omega=0$ without loss of generality.

\subsubsection{Lorentzian distribution, $n=1$.}
For a Lorentzian frequency distribution, $g_1(\omega)$, the first full analytic results have already been obtained by Ott and Antonsen in \cite{OttAntonsen2008}.
To evaluate the integral in \eqref{eq:zOA}, one applies Cauchy's residue theorem and integrates along the path encircling the lower complex half-plane.
For $n=1$ there is one pole and the governing dynamics obey \cite{OttAntonsen2008}
\begin{equation}\label{eq:unimodalLorentzian}
\dot{z}= - (\Delta + i\Omega) z + \tfrac{K}{2} \left[ z - |z|^2 z\right] \ .
\end{equation}
$\Omega$ denotes the center of the frequency distribution $g_1(\omega)$, which we consider to vanish though here we keep it for the sake of illustration.
In fact, we can rewrite \eqref{eq:unimodalLorentzian} more intuitively in Hopf normal form, that is, in the form of a Stuart Landau oscillator:
\begin{equation}\label{eq:unimodalLorentzianHNF}
\dot{z}= \left[(\tfrac{K}{2}-\Delta) + i\Omega \right] z - \tfrac{K}{2}  |z|^2 z \ .
\end{equation}
For positive coupling $K>0$, the order parameter $z$ undergoes a supercritical Hopf bifurcation at $K_c=\kappa_1 \Delta=2\Delta$, and for larger coupling $K>K_c$ there are stable limit cycle oscillations with frequency $\Omega$.

\subsubsection{Quartic distribution, $n=2$.}
For the quartic frequency distribution, $g_2(\omega)$, the Kuramoto order parameter $z=z_1+z_2$ comprises the dynamics of the pseudo order parameters $z_{1,2}$ with dynamics \cite{Skardal2018}
\begin{equation}\label{eq:unimodalQuartic}
\begin{aligned}
\dot{z}_1 &= -\frac{1-i}{\sqrt{2}}\Delta z_1 + \frac{K}{4} \left[ (1-i) (z_1+z_2) - 4\frac{z_1^\ast + z_2^\ast}{1-i} z_1^2 \right] \\
\dot{z}_2 &= -\frac{1+i}{\sqrt{2}}\Delta z_2 + \frac{K}{4} \left[ (1+i) (z_1+z_2) - 4\frac{z_1^\ast + z_2^\ast}{1+i} z_2^2 \right] .
\end{aligned}
\end{equation}
As this system is two-dimensional, it is convenient to rewrite it in the `new' variables $z$ and $m=z_1-z_2$, which yields
\begin{equation}\label{eq:unimodalQuartic2}
\begin{aligned}
\dot{z} &= \tfrac{K - \sqrt{2}\Delta}{2} z + i\tfrac{\Delta}{\sqrt{2}} m - \tfrac{K}{4} \left[ z^\ast (z^2 + m^2) + 2i|z|^2 m \right]\\
\dot{m} &= -\tfrac{\Delta}{\sqrt{2}} m + i\tfrac{\sqrt{2}\Delta-K}{2} z - \tfrac{K}{4} \left[ 2 |z|^2 m + iz^\ast (z^2 + m^2)  \right] .
\end{aligned}
\end{equation}
We are particularly interested in the transition from incoherence, $z=0$, to partial synchrony, $z>0$, which occurs for some critical coupling strength $K_c=\kappa_2 \Delta$.
When considering the dynamics around $z=0=z_1=z_2$, which also implies $m=0$, one can set $z_k\approx z/n=z/2$ for all $k=1,2$.
Then, we have $z_1+z_2=z$ as before, but $z_1-z_2\approx 0$.
In this small $z_k$-approximation the $z$-dynamics becomes
\begin{equation}\label{eq:unimodalQuarticHNF}
\dot{z} \approx \left[ \tfrac{K - \sqrt{2}\Delta}{2} + i\Omega \right] z - \tfrac{K}{4} |z|^2 z \ ,
\end{equation}
which is of similar character as the Hopf normal form \eqref{eq:unimodalLorentzianHNF} for the Lorentzian distribution.
Once the coupling $K$ exceeds $\sqrt{2}\Delta$, i.e. $\kappa_2=\sqrt{2}$, the incoherent state $z=0$ loses stability in a supercritical Hopf bifurcation and gives rise to stable limit cycle oscillations with frequency $\Omega$.

\subsubsection{Rational distributions with $n>2$.}
The above approach to determine the phase transition from incoherence to partial synchrony can be naturally extended to larger $n>2$.
Again, it is possible to rewrite the system $z_1,z_2,\dots, z_n$ exactly in terms of the Kuramoto order parameter $z=\sum_k z_k$ and another $n-1$  variables of the form $m_k:=z_1-z_k$.
Considering the dynamics around the incoherent solution $z=0=z_k$ for all $k$, such that $m_k=0$ for all $k$, we can use the approach that $z_k=z^0_k+\varepsilon z^1_k + \mathcal{O}(\varepsilon^2)$ with $z^0_k=z^0$ for all $k$.
We can thus  substitute $z_k=z/n + \mathcal{O}(\varepsilon)$ in the $z$-dynamics.
Upon summation of \eqref{eq:lowdimdynamics} and discarding all terms of order $\mathcal{O}(\varepsilon)$, we obtain
\begin{equation}\label{eq:unimodalCaseNApprox}
\begin{aligned}
\dot{z}&= \sum_{k=1}^n \dot{z}_k \approx \tfrac{i}{n}\Big( \sum_{k=1}^n \hat{w}_k \Big)^\ast z \ + \\
&\hspace{.4cm} \tfrac{K}{2} \left\lbrace 2\pi i  \Big(\sum_{k=1}^n r_{n,k} \Big)^\ast z - \tfrac{1}{n^2} \Big[2\pi i  \Big(\sum_{k=1}^n r_{n,k}\Big)^\ast \Big]^{-1}  |z|^2 z \right\rbrace.
\end{aligned}
\end{equation}
For the sums in \eqref{eq:unimodalCaseNApprox}, we have
\begin{equation}\label{eq:unimodalCaseNApproxHelpSums}
\begin{gathered}
\sum_{k=1}^n \hat{w}_k = \Delta  \sum_{k=1}^n \exp\big( -i\tfrac{2(k-1)+1}{2n}\pi\big) = -i \frac{\Delta}{\sin\left(\tfrac{\pi}{2n}\right)} \\
\sum_{k=1}^nr_{n,k} = \tfrac{-1}{2\pi} \sum_{k=1}^n \exp\big( -i\tfrac{2(k-1)+1}{2n}\pi \big) \sin\big( \tfrac{\pi}{2n} \big)  = \frac{-1}{2\pi i} 
\end{gathered}
\end{equation}
and thus find
\begin{equation}\label{eq:unimodalCaseNApproxHNF}
\dot{z} \approx \left[ - \frac{\Delta}{n \sin\left(\tfrac{\pi}{2n}\right)} + \frac{K}{2} + i \Omega \right] z - \frac{K}{2n^2} |z|^2 z \ .
\end{equation}
This approximation provides three crucial insights.
First, the incoherent solution $z=0$ loses stability at the critical coupling strength $K_c=\kappa_n \Delta = 2\Delta/\left[n\sin(\pi/2n)\right]$.
This value coincides also with Kuramoto's self-consistency argument in \cite{Kuramoto1984} for a unimodal frequency distribution $g(\omega)$.
According to Kuramoto, the critical coupling is $K^\ast=2/\left[\pi g(\Omega)\right]$ with $\Omega$ the center of symmetry of the frequency distribution.
In our case $\Omega=0$ and $g(0) = n\sin(\pi/2n)/ (\Delta \pi)$, such that $K^\ast = \kappa_n\Delta = K_c$.
Second, we can compute the critical coupling strength for the unimodal compact distribution in the limit $n\to\infty$. The scaling factor $\kappa_c=\kappa_\infty$ is given by
\begin{equation}
\kappa_c = \lim_{n\to\infty} \kappa_n = \lim_{n\to\infty} \frac{2}{n\sin\big(\tfrac{\pi}{2n}\big)} = \frac{4}{\pi} \ ,
\end{equation}
and the critical coupling strength $K_c=4\Delta/\pi$.
This result has already been found by Paz\'o in \cite{Pazo2005}.
Third, the Kuramoto order parameter undergoes a supercritical Hopf bifurcation for all $n=1,2,\dots$.
However, for $n\to \infty$ the cubic part of the Hopf normal form \eqref{eq:unimodalCaseNApproxHNF} vanishes and the bifurcation becomes degenerate.
This may already hint at the change of character of the phase transition from second to first order in the limit $n\to \infty$; see Fig.~\ref{fig:ExplosiveSync}.

\section{Collective dynamics for bimodal rational frequency distributions}
\label{sec:bimodal_rational}
\begin{figure}
\centerline{\includegraphics[width=1\columnwidth]{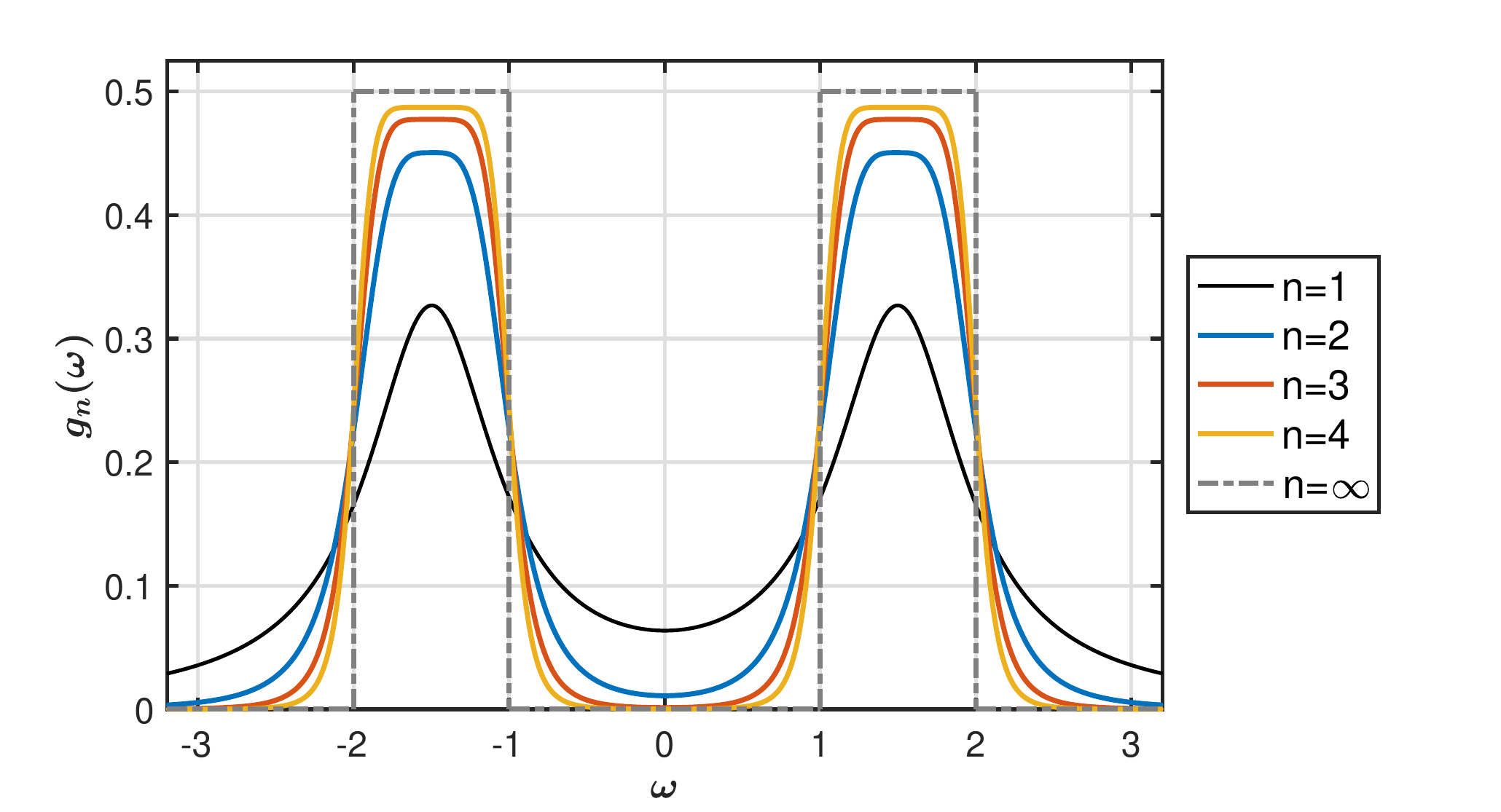}}
\caption{Flat bimodal frequency distributions. We plot the first four bimodal rational distributions $\tilde{g}^{(\Delta,\omega_0)}_n(\omega)$, $n=1,2,3,4$, and the bimodal compact distribution $\tilde{g}_c(\omega)$ as given in \eqref{eq:bimodalrationaldistribution} and \eqref{eq:bimodalcompactdistribution}, respectively, with $(\Delta,\omega_0)=(0.5,1.5)$.}
\label{fig:frequencies}
\end{figure}

To generalize the previous analytic approach to bimodal distributions, we consider bimodal frequency distributions that are symmetric around the origin $\Omega=0$.
We restrict them to be the sum of two symmetric rational frequency distributions of the form \eqref{eq:rationalfrequency},
\begin{align}
\tilde{g}_n(\omega)&=\tilde{g}^{(\Delta,\omega_0)}_n(\omega) = \tfrac{1}{2} \left( g^{(\Delta,-\omega_0)}_n(\omega) + g^{(\Delta,+\omega_0)}_n(\omega) \right)\nonumber\\
&=n \sin(\pi/2\pi) \frac{\Delta^{2n-1}}{2\pi} \label{eq:bimodalrationaldistribution}\\
&\hspace{.2cm} \cdot \left[ \frac{1}{(\omega-\omega_0)^{2n} + \Delta^{2n}} + \frac{1}{(\omega+\omega_0)^{2n} + \Delta^{2n}} \right] .\nonumber
\end{align}
In the limit $n\to\infty$, we retain the bimodal compact distribution
\begin{equation}\label{eq:bimodalcompactdistribution}
\tilde{g}_c(\omega) = \tfrac{1}{4\Delta} \big( \delta_{[-\Delta,\Delta]}(\omega-\omega_0) + \delta_{[-\Delta,\Delta]}(\omega+\omega_0) \big) \ ,
\end{equation}
where $\Delta\geq 0$ is the half-width of both blocks, which are centered around $\pm \omega_0$; $\delta_\mathcal{I}(\omega)$ denotes the Kronecker-$\delta$, which is $1$ if $\omega$ lies in the interval $\mathcal{I}$, and $0$ otherwise.
Fig.~\ref{fig:frequencies} shows the first four bimodal rational frequency distributions together with the bimodal compact distribution.

The Kuramoto model with bimodal rational frequency distributions can be fully analyzed along the OA ansatz, analogously to the previous section.
The system (\ref{eq:alphaOA}~\&~\ref{eq:zOA}) captures the order parameter dynamics also in the bimodal case.
To evaluate the integral in \eqref{eq:zOA}, we again dwell on Cauchy's residue theorem.
For this, we have to identify the poles $\tilde{w}_k$ of $\tilde{g}_n(\omega)$ and the respective residue values $\mathrm{Res}(\tilde{g}_n; \tilde{w}_k)$ to close the equation as for the dynamics \eqref{eq:lowdimdynamics}.
Due to the symmetric form of $\tilde{g}_n(\omega)$, there are $2n$ simple poles $\tilde{\omega}_k$ for each $n$ in the lower complex half-plane and can define $2n$ pseudo order parameters $z_k$, which results in a $2n$-dimensional system \eqref{eq:lowdimdynamics}.
The first $n$ poles are given by $\tilde{\omega}_k = \hat{\omega}_k - \omega_0$, $k=1,\dots,n$, and the latter $n$ by $\tilde{\omega}_k = \hat{\omega}_k + \omega_0$, $k=n+1,\dots,2n$.
Shifting the poles does not alter the residue values such that $\mathrm{Res}(\tilde{g}_n; \tilde{w}_k) =\mathrm{Res}(\tilde{g}_n; \tilde{w}_{n+k})= \mathrm{Res}({g}_n; \hat{w}_k)$ for $k=1,\dots,n$.
Next, we define left and right local order parameters as $z_l = \sum_{k=1}^n z_k$ and $z_r = \sum_{k=n+1}^{2n} z_k$, respectively.
Together they form the global order parameter \eqref{eq:orderparameter} as
\begin{equation}
z = \tfrac{1}{2} \big( z_l + z_r \big) \ .
\end{equation}
In consequence, the governing dynamics are given by
\begin{equation}\label{eq:bimodallowdimdynamics}
\begin{aligned}
\dot{z}_k = &\ i \tilde{\omega}_k^\ast z_k + \frac{K}{4} \left[ 2\pi i \big(\mathrm{Res}(\tilde g_n; \tilde{\omega}_k)\big)^\ast \Big( \sum_{j=1}^{m} z_j \Big) \right.\\
& \left.- \Big( \sum_{j=1}^{m} z_j^\ast \Big) \frac{z_k^2}{2\pi i \big(\mathrm{Res}(\tilde g_n; \tilde{\omega}_k)\big)^\ast} \right].
\end{aligned}
\end{equation}
Before returning to the dynamics of the Kuramoto model with a bimodal compact frequency distribution, we first analyze the system with bimodal rational distributions $\tilde{g}_n$ for small $n\in \mathbb{N}$.
The results for finite $n<\infty$ will naturally extend to $n\to\infty$, i.e, to the bimodal (discontinuous) compact distribution.
As the number of differential equations is $2n$, the complex dynamics become intractable for large $n$.
Yet, as we will show below, \eqref{eq:bimodallowdimdynamics} provides invaluable information about the bimodal compact Kuramoto model.
We split the remainder of the section into a part about the different possible bifurcations off the incoherent solution and a part about the complete bifurcation diagram.
In each part, we increase the parameter $n\geq 1$ so that the link to the compact distribution, $n\to\infty$, becomes apparent.


\subsection{Linear stability of the incoherent solution}

\subsubsection{Bimodal Lorentzian distribution, $n=1$}
The Kuramoto model with a bimodal Lorentzian frequency distribution has been extensively analyzed by Martens and co-workers in \cite{MartensExactResults2009}, see also \cite{PiDeDa2016}.
For a non-vanishing distance $\omega_0 > 0$ between the two (symmetric) peaks of the distribution $\tilde{g}_1(\omega)$, the local order parameters $z_r$ and $z_l$ defined above gain some illustrative power.
Their dynamics are given by \eqref{eq:bimodallowdimdynamics} for $n=1$ \cite{MartensExactResults2009}.
Note that $\omega_0 > 0$ results in a shift of the poles $\hat{w}_k$ towards $\tilde{w}_k = \hat{w}_k \pm \omega_0$, but leaves the residue values identical.
Rewriting the local order parameter dynamics in form of two Stuart-Landau oscillators, one finds
\begin{equation}
\begin{aligned}
\dot{z}_l &= \big[ \big( \tfrac{K}{4} - \Delta\big) - i\omega_0 \big] z_l - \tfrac{K}{4} |z_l|^2 z_l +  \tfrac{K}{4} \left\lbrace z_r - z_r^\ast z_l^2  \right\rbrace \\
\dot{z}_r &= \big[ \big( \tfrac{K}{4} - \Delta\big) + i\omega_0 \big] z_r - \tfrac{K}{4} |z_r|^2 z_r +  \tfrac{K}{4} \left\lbrace z_l - z_l^\ast z_r^2  \right\rbrace \ .
\end{aligned}
\end{equation}
The curly brackets denote coupling terms (linear and nonlinear) to the respective other local order parameter.
As we are interested in the critical behavior close to the onset of synchronization, the real parts of the linear coefficients can be considered small.
The coupling strength will be of order $\mathcal{O}(\Delta)$.
Unfortunately, the theory of weakly coupled oscillators may no longer apply in this case.
It is hence inevitable to consider the full dynamics in two complex dimensions or in four real dimensions.
Linearizing the dynamics around $z_l=0=z_r$ yields the degenerate eigenvalues \cite{MartensExactResults2009,PiDeDa2016}
\begin{equation}
\begin{aligned}
\lambda_1 &= \lambda_2 = \tfrac{K}{4} - \Delta + \tfrac{K}{4}\sqrt{1-\big(\sfrac{4\omega_0}{K}\big)^2}\\
\lambda_3 &= \lambda_4 = \tfrac{K}{4} - \Delta - \tfrac{K}{4}\sqrt{1-\big(\sfrac{4\omega_0}{K}\big)^2} 
\end{aligned}
\end{equation}
and the resulting stability boundaries, at which the incoherent solution becomes unstable, are
\begin{equation}
\Delta = \frac{K}{4} + \begin{cases} \tfrac{K}{4} \sqrt{1-\big(\sfrac{4\omega_0}{K}\big)^2} \quad &\text{for } 0 \leq \omega_0 \leq \sfrac{K}{4}\\
0 & \text{otherwise}. \end{cases}
\end{equation}
The upper solution defines a transcritical bifurcation, whereas a supercritical Hopf bifurcation occurs for $\omega_0 \geq K/4$.

\subsubsection{Bimodal quartic distribution, $n=2$}
For the quartic frequency distribution, each local order parameter comprises two pseudo order parameters, $z_l = z_1+z_2$ and $z_r=z_1+z_2$.
While the first two order parameters $z_1,z_2$ follow the dynamics \eqref{eq:unimodalQuartic} with additional linear parts $-i\omega_0 z_1$ and $-i\omega_0 z_2$, respectively, the latter two have additional linear parts $+i\omega_0 z_3$ and $+i\omega_0 z_4$:
\begin{widetext}
\begin{equation}\label{eq:bimodalQuartic}
\begin{aligned}
\dot{z}_1 &=  -\left[ \frac{1-i}{\sqrt{2}}\Delta  +i\omega_0 \right] z_1 + \frac{K}{8} \left[ (1-i) (z_1 + z_2+z_3 + z_4) - 4\frac{z_1^\ast + z_2^\ast +z_3^\ast + z_4^\ast}{1-i} z_1^2 \right]\\
\dot{z}_2 &=  -\left[ \frac{1+i}{\sqrt{2}}\Delta  +i\omega_0 \right] z_2 + \frac{K}{8} \left[ (1+i) (z_1 + z_2+z_3 + z_4) - 4\frac{z_1^\ast + z_2^\ast +z_3^\ast + z_4^\ast}{1+i} z_2^2 \right]\\
\dot{z}_3 &=  -\left[ \frac{1-i}{\sqrt{2}}\Delta  -i\omega_0 \right] z_3 + \frac{K}{8} \left[ (1-i) (z_1 + z_2+z_3 + z_4) - 4\frac{z_1^\ast + z_2^\ast +z_3^\ast + z_4^\ast}{1-i} z_3^2 \right]\\
\dot{z}_4 &=  -\left[ \frac{1+i}{\sqrt{2}}\Delta  -i\omega_0 \right] z_4 + \frac{K}{8} \left[ (1+i) (z_1 + z_2+z_3 + z_4) - 4\frac{z_1^\ast + z_2^\ast +z_3^\ast + z_4^\ast}{1+i} z_4^2 \right]
\end{aligned}
\end{equation}
\end{widetext}
Similar to \eqref{eq:unimodalQuartic2} one can rewrite the dynamics in the variables $z_l,z_r,m_l,m_r$ with $m_l = z_1 - z_2$ and $m_r=z_3-z_4$.
As the linear dynamics of $m_l$ and $m_r$ are independent of the coupling $K$, one can assume that $m_l=m_r=0$ for small $|z_l|, |z_r| \ll 1$.
Setting $z_1=z_2=z_l/2$ and $z_3=z_4=z_r/2$ yields the approximate dynamics
\begin{equation}
\begin{aligned}
\dot{z}_l &\approx \big[\big(\tfrac{K}{4}- \tfrac{\Delta}{\sqrt{2}} \big) -i\omega_0\big] z_l + \tfrac{K}{8} |z_l|^2z_l + \tfrac{K}{8} \left\lbrace 2 z_r - z_r^\ast z_l^2 \right\rbrace \\
\dot{z}_r &\approx \big[\big(\tfrac{K}{4}- \tfrac{\Delta}{\sqrt{2}} \big) +i\omega_0\big] z_r + \tfrac{K}{8} |z_r|^2z_r + \tfrac{K}{8} \left\lbrace 2 z_l - z_l^\ast z_r^2 \right\rbrace
\end{aligned}
\end{equation}
Similar to the bimodal Lorentzian case, one can solve for the eigenvalues of the Jacobian corresponding to the linearized dynamics around $z_r=z_l=0$.
This reveals the bifurcation boundaries:
\begin{equation}\label{eq:bimodalQuarticBifBoundaryApprox}
\Delta = \frac{K}{2\sqrt{2}} + \begin{cases} \tfrac{K}{2\sqrt{2}} \sqrt{1-\big(\tfrac{2\sqrt{2}\omega_0}{K}\big)^2} \quad &\text{for } 0 \leq \omega_0 \leq \tfrac{K}{2\sqrt{2}}\\
0 & \text{otherwise}. \end{cases}
\end{equation}
The line $\Delta = K/2\sqrt{2}$ denotes a Hopf bifurcation for $\omega_0 \geq K/2\sqrt{2}$.
Remarkably, when investigating the full system \eqref{eq:bimodalQuartic} numerically, the collective dynamics displays the predicted Hopf bifurcation at the critical coupling strength $K_c=2\sqrt{2}\Delta = 2\kappa_2\Delta$.
However, along the arch-shaped transcritical bifurcation boundary the collective dynamics are partially synchronized with $|z_l|, |z_r| \gg 0$.
We therefore conclude that our approach holds for the emergence of collective oscillations through the Hopf bifurcation, but it is not valid to predict any other bifurcations.
With the simplifying assumptions above we reduced the actual $8$-dimensional system to only $4$ (real) dimensions in \eqref{eq:bimodalQuarticBifBoundaryApprox}.
Due to this reduction in complexity we ignore the correct synchronization effects on the left of the diagonal $\Delta=\omega_0$.

A possible explanation why this approach fails to predict the transcritical bifurcation boundary may be given by the distribution of poles $\tilde{w}_k$.
Reconsidering \eqref{eq:bimodalQuartic}, we focus on the uncoupled linear part of the dynamics.
While the real part, $-\Delta/\sqrt{2}$, is the same for all $z_k$, $k=1,\dots,4$, the position of the poles directly influences the imaginary parts.
The dynamics have imaginary linear parts $\pm\omega_0 \pm i\Delta/\sqrt{2}$.
In line with the approach above we group those dynamics together with the same sign of $\omega_0$.
For a given value of $\omega_0>0$, at a critical half-width $\Delta_c=\Delta_c(\omega_0)$ the linear parts of $z_1$ and $z_4$ are closer to each other than those of $z_1$ and $z_2$, which are combined as the left local order parameter $z_l$.
Hence, the motivation to fix $z_l=z_1+z_2$ becomes questionable, and one may rather consider three (clustered and interacting) order parameters $z_A=z_2$, $z_B=(z_1+z_4)/2$, and $z_C=z_3$.
The analysis of this system in three complex variables becomes almost as intricate as the one of the full system, so that we will rely on a semi-analytic analysis in the following to detect the transcritical bifurcation boundary.

\subsubsection{Hopf bifurcation for bimodal rational distributions, $n\geq1$.}

Although our approach is not appropriate to derive the transcritical bifurcation, we can still use it to establish the Hopf bifurcation boundary.
This can naturally be extended to larger $n>2$, rendering the derivation general.
In fact, the corresponding Hopf normal form close to the bifurcation point will be derived for the global order parameter $z=(z_l+z_r)/2$.
We follow a similar reasoning as around \eqref{eq:unimodalCaseNApprox} -- \eqref{eq:unimodalCaseNApproxHNF}.
Capitalizing on the symmetry of the bimodal frequency distribution $\tilde g_n(\omega)$ (such that the $\pm \omega_0$ terms will cancel), we can sum the dynamics \eqref{eq:bimodallowdimdynamics} for $k=1,\dots,2n$ and employ the identities \eqref{eq:unimodalCaseNApproxHelpSums}.
This results in an approximate expression, similar to \eqref{eq:unimodalCaseNApproxHNF} in the unimodal case,
\begin{equation}\label{eq:bimodalHopf}
\dot{z} \approx \left[ \frac{-\Delta}{n\sin\big(\tfrac{\pi}{2n}\big)} + \frac{K}{4} + i\Omega \right] z - \frac{K}{2n^2} |z|^2 z \ .
\end{equation}
This approximation is only valid for small $|z_k| \ll 1$, $k=1,\dots,2n$.
In the vicinity of the incoherent solution $z=0$, however, it predicts the Hopf bifurcation at critical coupling strength $K_H=2\kappa_n \Delta$ with scaling factors 
\begin{equation}\label{eq:couplingscalingfactor}
\kappa_n = \frac{2}{n \sin\big(\tfrac{\pi}{2n}\big)} \ .
\end{equation}	

\subsubsection{Transcritical bifurcation for bimodal rational distributions, $n\geq1$.}

\begin{figure}[h!]
\centerline{\includegraphics[width=.85\columnwidth]{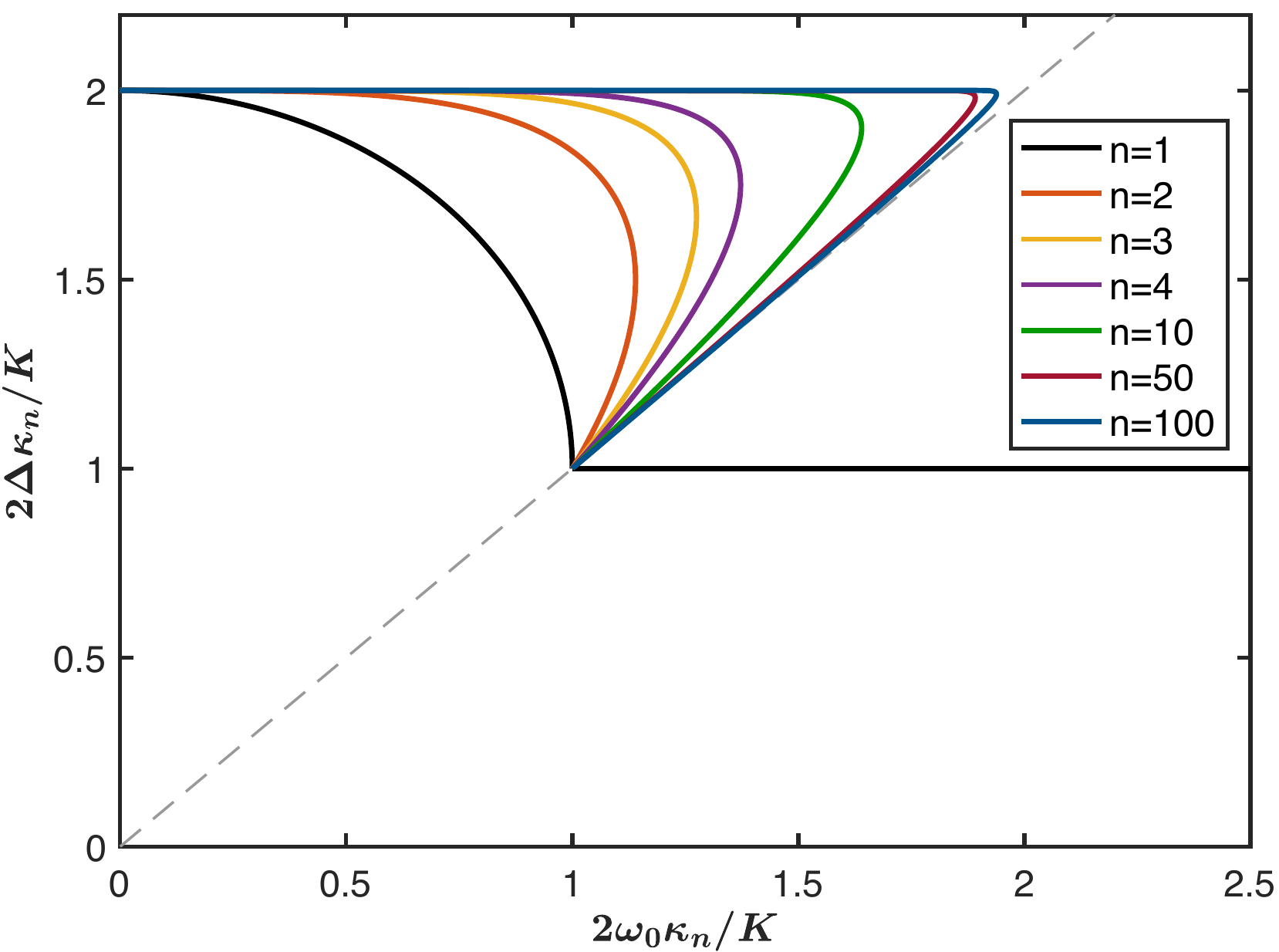}}
\caption{Stability boundaries of the incoherent solution for bimodal rational frequency distributions $\tilde g_n(\omega)$.
The bifurcation parameters $\Delta$ and $\omega_0$ are rescaled such that the Hopf bifurcation occurs at $2\Delta\kappa_n/K=1$ with scaling factors are given by $\kappa_n = 2/\left[ n \sin(\pi/2n)\right]$.}
\label{fig:stabilityboundaries}
\end{figure}

By exploiting the spectrum of the Jacobian numerically, one can determine the bifurcation boundaries of the incoherent solution, $z=z_k=0$ for all $k=1,\dots,2n$, for any bimodal rational frequency distribution $\tilde g_n(\omega)$ with $n\geq 1$.
The results are depicted in Fig.~\ref{fig:stabilityboundaries}.
The straight line with $2\Delta\kappa_n/K=1$ starting at $2\omega_0\kappa_n/K=1$ is the Hopf bifurcation identified above.
The bent curve with $2\Delta\kappa_n/K>1$, by contrast, is of transcritical type.
The transcritical bifurcation occurs for larger values of $\Delta$ and $\omega$ than those given by the arch-shaped bifurcation curve as predicted through \eqref{eq:bimodalQuarticBifBoundaryApprox}.
We would like to note that the distinction of the bifurcations off the incoherent solution is based on the dynamics observed for the global Kuramoto order parameter $z$ and the local order parameters, $z_l, z_r$, respectively. 
It is true that we consider the full dynamics of all $z_k$ to determine the spectrum of the Jacobian.
Yet, the particular structure of the poles $\tilde\omega_k$ around the (shifted) unit circle(s) opposes a certain degeneracy of the bifurcations, so that we characterize the different bifurcations according to the global behavior for the sake of conciseness.

\subsubsection{A note on critical coupling and Kuramoto's conjecture}
As outlined above, the critical coupling strength $K_H$ for the Hopf bifurcation coincides with the onset of synchrony in the unimodal network, $K_H= \kappa_n \Delta$ with $\kappa_n$ given by \eqref{eq:couplingscalingfactor}.
This result is insofar remarkable that $K_H$ coincides with $K_c=2/\left[ \pi \tilde g_n(\Omega) \right]$, where $\Omega=0$ because the natural frequency distributions $g_n$ are symmetric about the origin.
As mentioned earlier, $K_c$ was found to be exactly the critical coupling strength in Kuramoto's original self-consistency analysis \cite{Kuramoto1984}.  His result, however, was only valid for unimodal frequency distributions $g(\omega)$.
When the frequency distribution $g(\omega)$ is concave, i.e. $g''(0) > 0$, Kuramoto conjectured that there must be a different (smaller) coupling strength $\tilde{K}_c=2/\left[ \pi g(\omega_1) \right]$ with $\omega_1 \neq 0$.
Then, there must be at least two peaks, as, e.g., in our bimodal case.
$\tilde{K}_c$ denotes then the onset of nucleation around each of the two peaks.
Two giant clusters form that oscillate each at their own frequency.
More formally, we consider the local order parameters in polar coordinates, $z_l = R_l \mathrm{e}^{i\phi_l}, z_r = R_r \mathrm{e}^{i\phi_r}$.
The nucleation process described above means that both $\dot{R}_l=0=\dot{R}_r$ with $R_l,R_r>0$, but their phase difference $\Psi =\phi_l - \phi_r$ is non-constant, $\dot{\Psi}\neq 0$.

In the case of our bimodal rational frequency distributions $\tilde g_n(\omega)$, there exists a value $\omega_1 \neq 0$ other than the axis of symmetry, $\Omega=0$, at which global oscillations emerge, demarcating the offset of partial synchrony.
This is particularly true for the bimodal Lorentzian distribution, $n=1$, and will also hold for small $n>1$.
In the limit $n\to \infty$, however, the transcritical bifurcation boundary will coincide with the diagonal $\Delta=\omega_0$, which presents a natural boundary for the bimodality of the frequency distributions $\tilde g_n$, $n=1,2,\dots, \infty$.
Beyond the diagonal, all the frequency distributions are no longer bimodal, that is, they are no longer concave. 
As can be seen in Fig.~\ref{fig:stabilityboundaries}, the transcritical bifurcation boundary converges to the diagonal for $n\to \infty$.
Hence, the compact bimodal distribution $g_c=g_{n\to\infty}$ is only bimodal in the parameter region below the diagonal.
Here, the smallest coupling value for the onset of synchronization is exactly the Hopf point $K_H=2/\left[ \pi \tilde g_n(0) \right]$.
We therefore conclude that for a symmetric, bimodal compact frequency distribution the critical coupling coincides with the Kuramoto's original formula for the onset of synchronization in a unimodal network.

\begin{figure}[htb!]
\centerline{\includegraphics[width=.85\columnwidth]{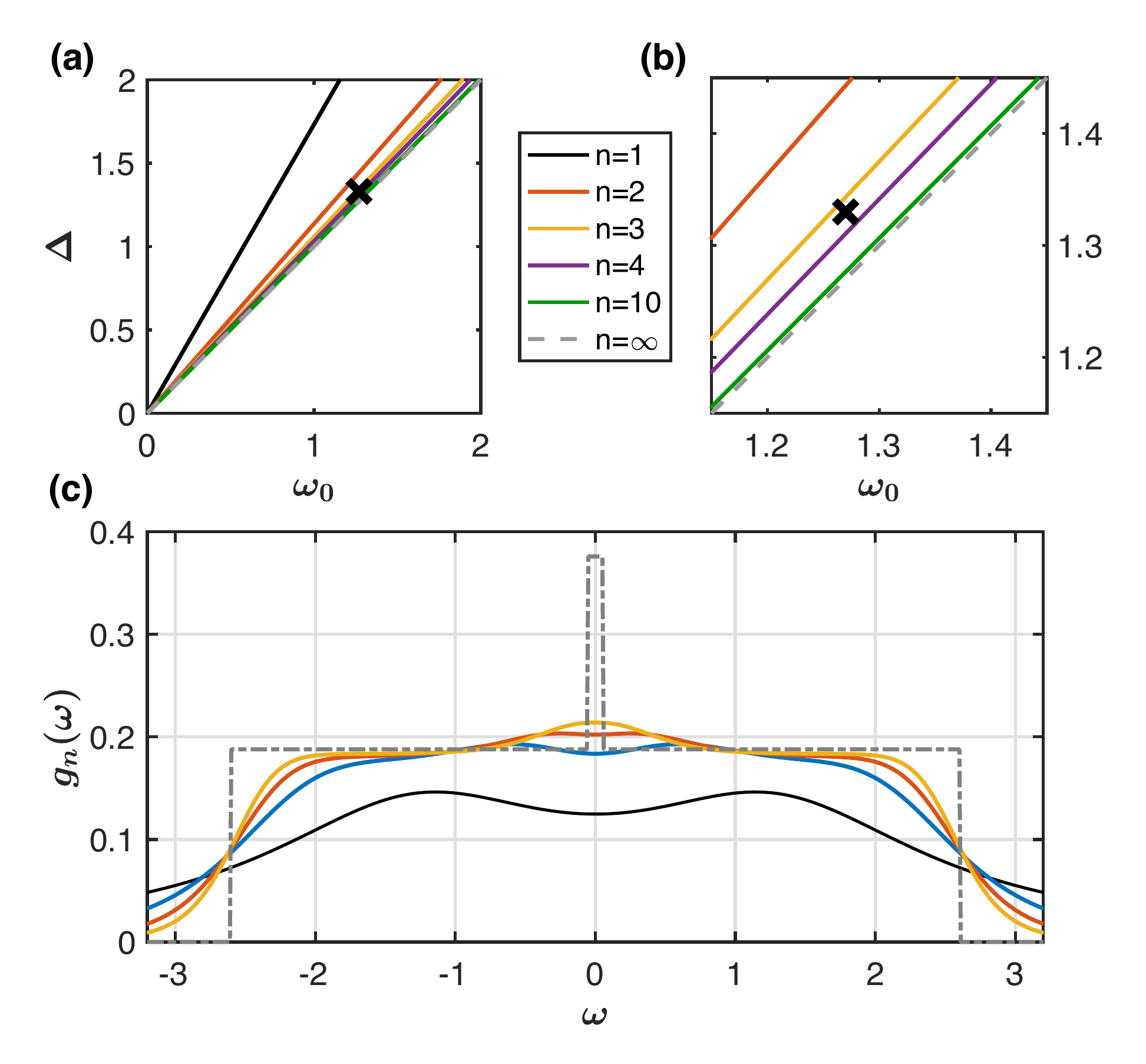}}
\caption{(a) Regions of bimodality for $\tilde{g}_n(\omega)$ are on the right of the colored curves. For increasing $n$ the straight lines converge to the diagonal $\Delta=\omega_0$, which defines the natural boundary between uni- and bimodality of the bimodal compact distribution $\tilde{g}_c(\omega)$.
(b) Zoom of (a) around the black cross at parameter values $(\Delta, \omega_0)=(1.33,1.27)$.
(c) Frequency distributions at the edge of bimodality at the black cross. While $\tilde{g}_n(\omega)$ is bimodal for small $n=1,2,3$, it becomes unimodal for $n > 3$. The bimodal compact distribution $\tilde{g}_c(\omega)$ shows a clear peak at the center of symmetry, $\Omega=0$.}
\label{fig:bimodregions}
\end{figure}

In order to illustrate the change from uni- to bimodality, we show the boundaries of bimodality in the $\omega_0-\Delta$ plane for frequency distributions $\tilde g_n$ in Fig.~\ref{fig:bimodregions}.
On the left of the colored curves $\tilde g_n$ is unimodal, while it becomes bimodal on the right.
The bimodal region for the Lorentzian case $n=1$ is largest. 
For larger $n$ the boundaries converge towards the diagonal $\Delta=\omega_0$.
For a particular parameter pair $(\Delta, \omega_0) =(1.33,1.27)$ slightly to the left of the diagonal (see black cross), we plot some rational distributions next to the compact one in Fig.~\ref{fig:bimodregions}~(c).
While the Lorentzian ($n=1$), the quartic ($n=2$) and the sextic ($n=3$) are still bimodal, already the octic distribution ($n=4$) becomes concave at the center of symmetry $\Omega=0$.
Clearly, the compact distribution has become unimodal. 

\subsection{Complete bifurcation diagrams for $n<\infty$}

\subsubsection{Bimodal Lorentzian distribution, $n=1$}
\begin{figure}[!b]
\centerline{\includegraphics[width=.95\columnwidth]{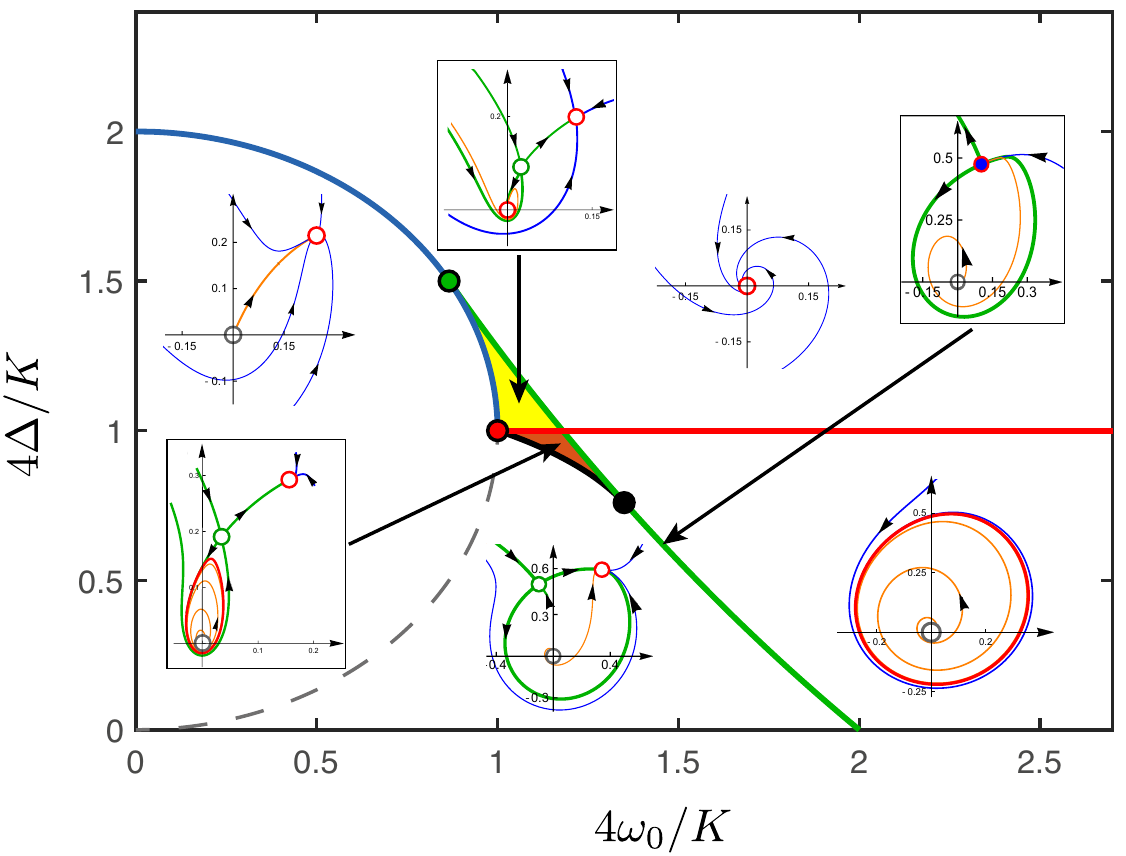}}
\caption{Bifurcation diagram for the bimodal Lorentzian distribution $\tilde g_1(\omega)$.
The solid curves depict supercritical Hopf (red), transcritical (blue), saddle-node (green) and homoclinic (black) bifurcations.
The red filled dot is a Bogdanov-Takens bifurcation. The black dot (saddle-node-loop bifurcation) delimits the SNIC bifurcation from the (upper) saddle-node curve (both green solid).
The colored regions are areas of bistability: either two stable fixed-points coexist (yellow), or a stable fixed-point and a stable limit-cycle (brown).
Insets show typical behavior of the order parameter dynamics in the corresponding parameter regions, see the main text for a description.}
\label{fig:martensAnalysis}
\end{figure}

As said, the first complete analytic picture of the dynamical regimes of the Kuramoto model with a bimodal frequency distributions has been provided by Martens and co-workers in \cite{MartensExactResults2009}.
For the sake of completeness, we briefly revisit the various transitions between the dynamical regimes, depicted in Fig.~\ref{fig:martensAnalysis}.
According to the stability analysis of the incoherent state $z \equiv 0$, the Hopf and transcritical bifurcations form an integral part of the bifurcation diagram. 
It is possible to determine the saddle-node curve analytically, which starts at $(\omega_0,\Delta) = (K/2,0)$ and adapts asymptotically to the transcritical arch at $(\sqrt{3}K/8,3K/8)$; see \cite{MartensExactResults2009} for more details and the exact expression.
For small $\Delta$, the saddle-node bifurcation occurs directly on the limit cycle, giving rise to a SNIC bifurcation (a saddle-node bifurcation occurs on the stable limit cycle). 
For large $\Delta$, on the other hand, we find regions of bistability as the saddle-node bifurcation occurs off the limit-cycle (below the Hopf bifurcation line), and off the incoherent solution (above).
Furthermore, the intersection of the Hopf and transcritical bifurcation boundaries is a Bogdanov-Takens bifurcation (red filled dot), which is a bifurcation of co-dimension 2.
From here a homoclinic bifurcation curve emerges and adapts to the saddle-node curve in a saddle-node-loop bifurcation (black dot).
Below this point, the saddle-node curve becomes the SNIC curve.

The insets in Fig.~\ref{fig:martensAnalysis} show characteristic dynamical behavior in each of the parameters regions.
Capitalizing on the symmetry of the bimodal distribution function, we can assume that the local order parameters $z_l, z_r$ have the same absolute value $|z_l|^2=|z_r|^2=q$ and only their respective angles, $\phi_l \neq \phi_r$ differ.
The insets thus depict the dynamics of $(q,\phi_l-\phi_r)$ transformed in Euclidean coordinates.
Red filled dots and red loops denote stable fixed points and limit cycles, respectively. The open dots are unstable (grey) and saddle (green) fixed points.
As we will show in the following, the same bifurcation structure will be maintained for larger $n>1$.

\subsubsection{Bimodal rational distributions, $n=2, 3, 4$}
\begin{figure*}[ht]
  \centering
  \setbox1=\hbox{\includegraphics[width=.6\columnwidth]{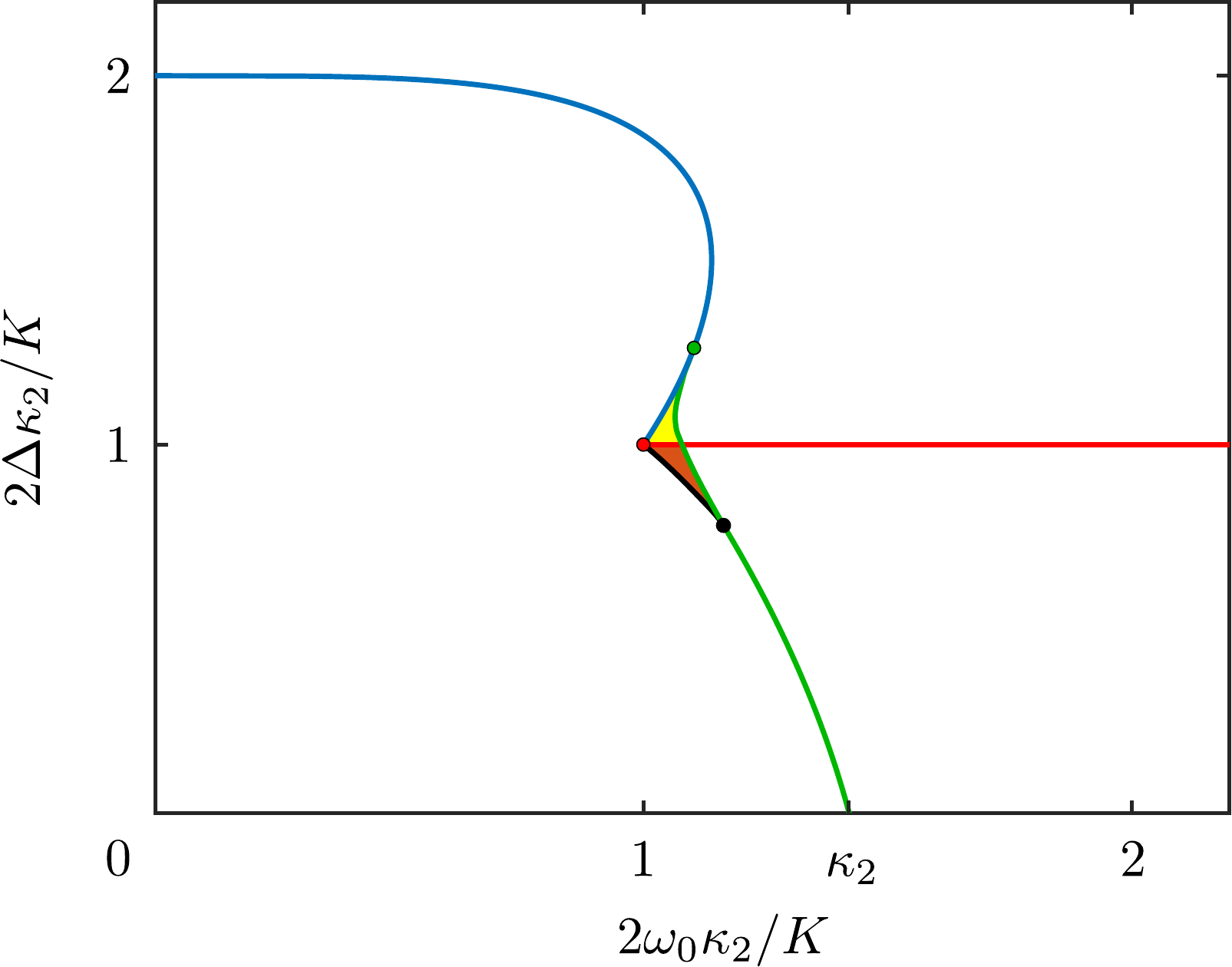}}
 \includegraphics[width=.6\columnwidth]{quarticBifurcationDiagram1}\llap{\makebox[.825\wd1][l]{\raisebox{.85cm}{\frame{\includegraphics[height=1.7cm]{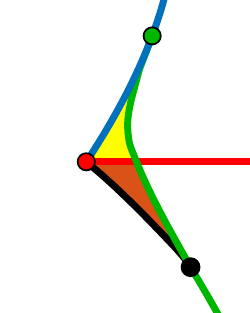}}}}}
\hspace{2em}
\includegraphics[width=.6\columnwidth]{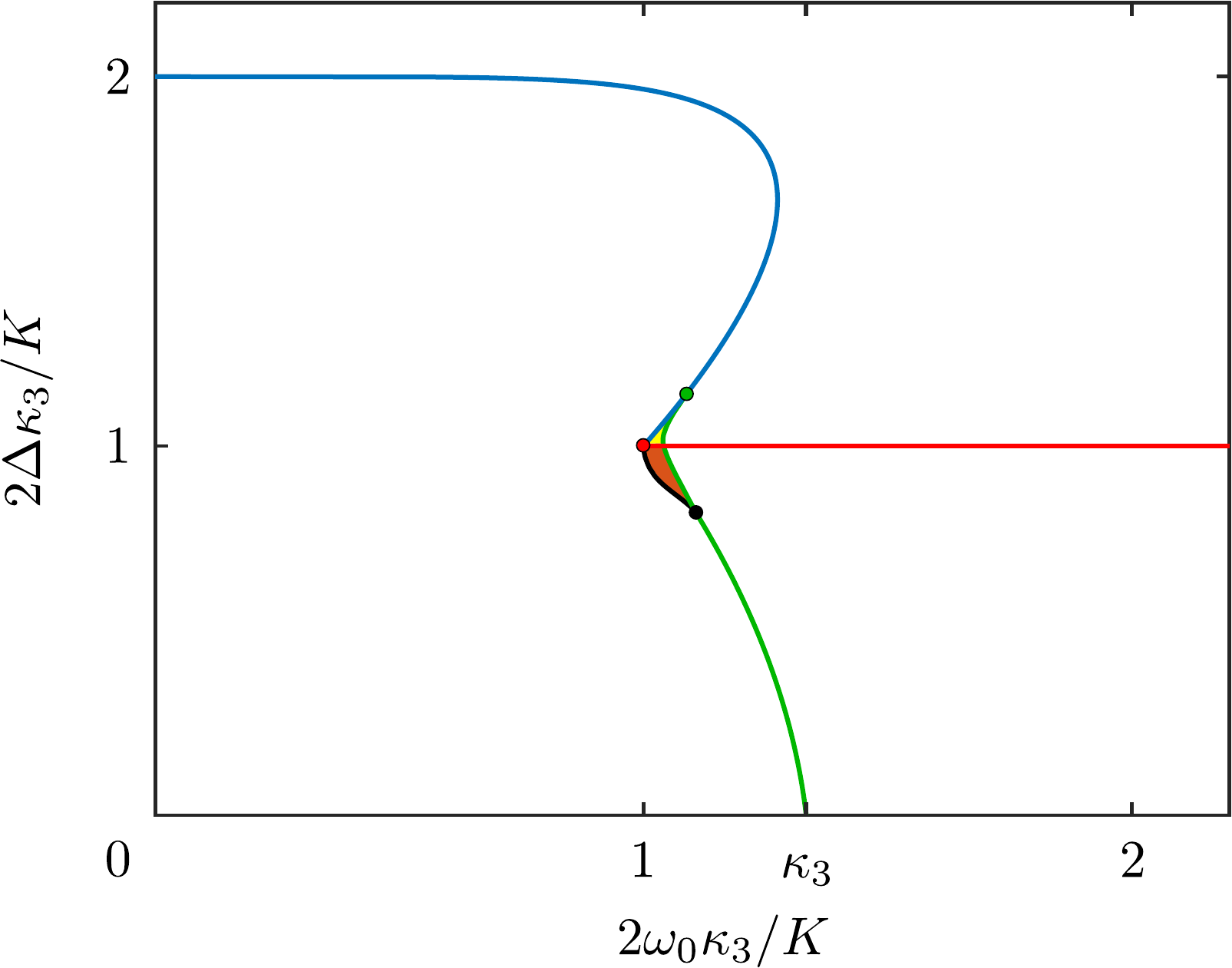}\llap{\makebox[.825\wd1][l]{\raisebox{.85cm}{\frame{\includegraphics[height=1.7cm]{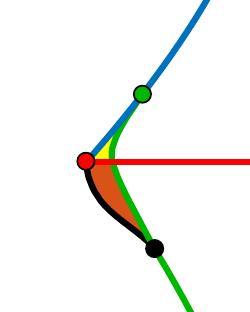}}}}}
\hspace{2em}
\includegraphics[width=.6\columnwidth]{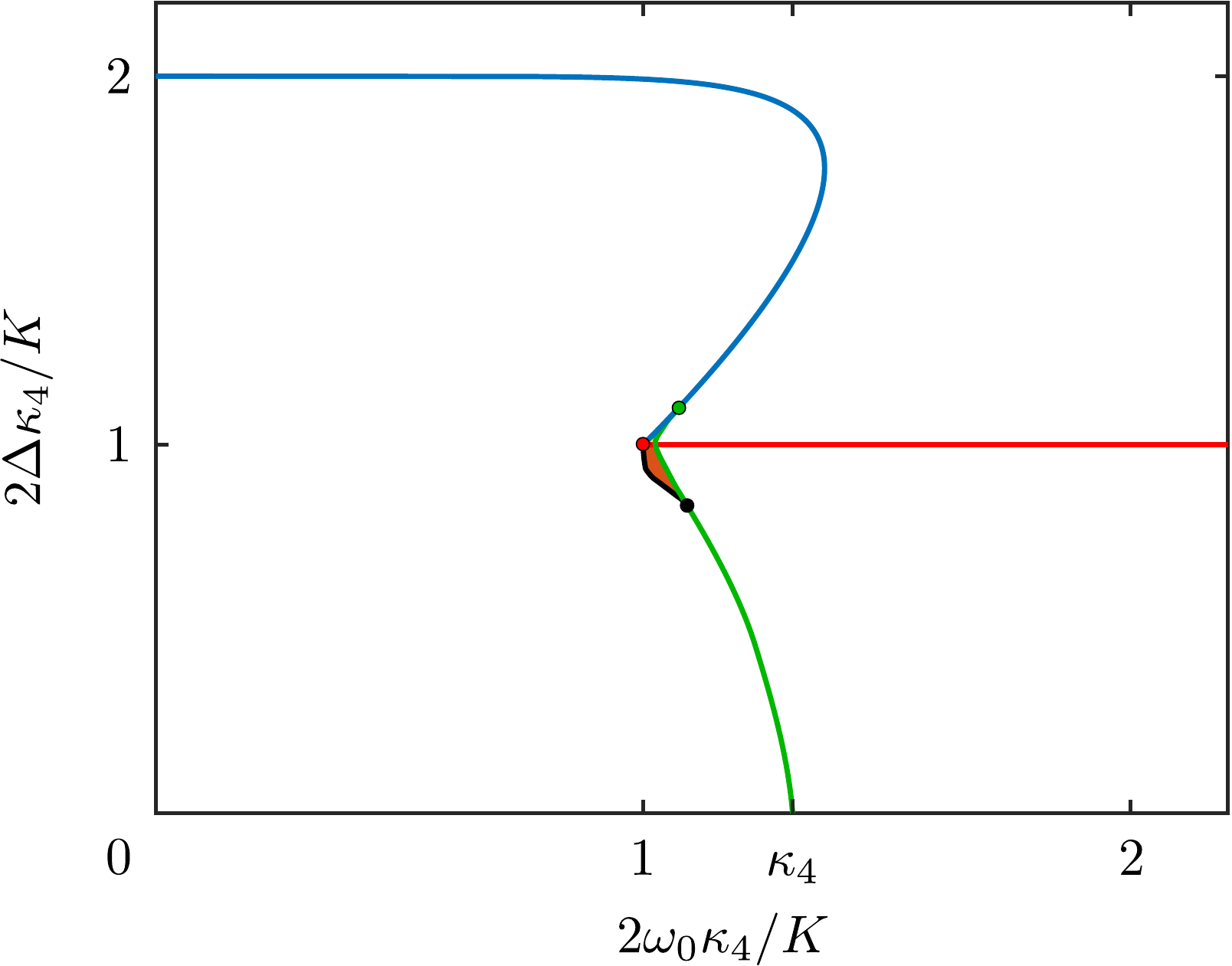}\llap{\makebox[.825\wd1][l]{\raisebox{.85cm}{\frame{\includegraphics[height=1.7cm]{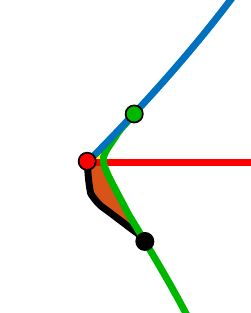}}}}}
\caption{Bifurcation diagrams for bimodal quartic, sextic and octic frequency distributions $n=2,3,4$.
The insets show a zoom into the bistability area with $\tilde{\omega}_0 \in [0.86,1.26]$ and $\tilde{\Delta} \in [0.68,1.34]$ with $\tilde{\omega}_0=2\omega_0\kappa_n/K$ and $\tilde{\Delta}=2\Delta\kappa_n/K$.
Color coding is the same as in Fig.~\ref{fig:martensAnalysis}.}
\label{fig:bifdiagramsN234}
\end{figure*}

Equipped with the results of the ``simplest'' bimodal rational frequency distribution, $\tilde g_1(\omega)$, we can also investigate the collective dynamics of the bimodal quartic ($n=2$), sextic ($n=3$) and octic ($n=4$) distributions.
We stick to the parameter scaling as before with scaling factors $\kappa_n$.
In Fig.~\ref{fig:bifdiagramsN234} the different bifurcation diagrams are summarized.
The insets show a zoom (with the same factor) into the region of bistability around the Bogdanov-Takens point (red).
Not only does the bistability region shrink for larger $n$, that is, the black (SNL bifurcation) and green dots move closer to each other.
But also the shape of the brownish region changes; in this area a partially synchronized state coexists with stable limit-cycle oscillations.
Another intriguing observation is the bending of the saddle-node curve (green) compared to the transcritical bifurcation boundary (blue).
Due to the scaling, the starting point of the saddle-node curve (more precisely, it is a SNIC bifurcation up to the SNL bifurcation point) always appears at $(\Delta,\omega_0)=(0,K/2)$.
This result is not surprising as in the limit $\Delta \to 0$ the bimodal distribution features two $\delta$-peaks at $\omega=\pm\omega_0$, so that the whole network consists of two symmetric oscillator populations each with identical frequencies.
For already small coupling strengths both populations synchronize completely and we can reduce the bimodal network to two single oscillators, for which the transition to mutual synchronization is well known to appear at the critical mutual coupling $K=2\omega_0$.
While $\kappa_4$ is already close to $\kappa_\infty=\kappa_c=4/\pi$, the offset of the saddle-node curve will not change much in the limit $n\to\infty$.
On the other hand, we expect a qualitative change of the bistability region (yellow and brown) as it already shrinks substantially for small but increasing $n>1$.
The shrinking of the yellow region is particularly dominant compared to the brown one, so that in case of the compact bimodal distribution the green dot is expected to coincide with the Bogdanov-Takens point, whereas there may remain a finite area in which stable limit-cycle oscillations coexist with a stable fixed-point solution.
The disappearance of the the yellow region of bistability may also be anticipated as the transcritical bifurcation boundary (blue curve) adapts asymptotically to the diagonal $\Delta=\omega_0$.
The following section will shed light on the actual dynamics and also answer the question whether bimodal compact (discontinuous) distributions still exhibit regions of bistability, or whether no bistable regions survive and the system therefore becomes less complex than those systems with bimodal frequency rational distributions with finite $n < \infty$.

\section{Collective behavior for bimodal compact frequency distribution}
\label{sec:bimodal_compact}
As `simple' as the (bimodal) compact distribution can be described, as difficult becomes the mathematical analysis of the corresponding collective dynamics.
The discontinuities at the edges of the plateaus do not allow for an analytic continuation of the compact distribution in the complex plane.
This, however, is a necessary requirement for the OA ansatz to be applied.

The introduction of the family of rational distribution functions \eqref{eq:rationalfrequency} seemed as a promising loophole as those distributions $g_n(\omega) (\tilde{g}_n(\omega))$ converge towards the discontinuous compact distribution $g_c(\omega) (\tilde{g}_c(\omega))$ in the limit $n\to\infty$.
Unfortunately, the dynamical system that describes the presumably low-dimensional time-asymptotic behavior of the network's order parameter is $n (2n)$ dimensional.
In the limit of the compact distribution the governing system of ordinary differential equations becomes infinite-dimensional, and hence no longer tractable.

Before resorting to numerical simulations in the last instance, we first approach the Kuramoto model with bimodal compact frequency distribution with the self-consistency argument as reappraised in \textit{Section}~\ref{subsec:self-cons_general}.

\subsection{Self-consistency approach}\label{subsec:self-cons_bimod}
Analogously to Paz\'o's analytic treatment of the unimodal compact distribution in \cite{Pazo2005}, we can apply the same argument to the bimodal compact distribution.
Even if this approach does not reveal the full bifurcation diagram, it may give important insight into the stability boundaries of the phase locked solutions.
\subsubsection{Incoherent solution}
We consider the stability of the incoherent solution $|z|=R=0$.
Near criticality we can assume that $R$ is small so that the right-hand side of the self-consistency equation \eqref{eq:self-consK} can be expanded in powers of $R$ while capitalizing on the symmetry of $\tilde{g}_c(\omega)$ around $\Omega=0$:
\begin{equation}\label{eq:bimodal_selfcons_approx}
1= \frac{\pi}{2} K \tilde{g}_c(0) + \frac{\pi}{16} K^3 R^2 \tilde{g}_c''(0) + \mathcal{O}(R^3) \ .
\end{equation}
The special form of the bimodal compact distribution $\tilde{g}_c(\omega)$ forces all derivatives to be zero except for at the discontinuities.
Moreover, $\tilde{g}_c(0)$ is zero if $\Delta < \omega_0$ such that \eqref{eq:bimodal_selfcons_approx} can never be fulfilled.
Still, on the left of the diagonal $\Delta=\omega_0$ we retrieve the critical coupling strengths $K_c = 4\Delta/\pi$ for $\Delta>\omega_0$ and $K_c=8\Delta/\pi$ on the diagonal.
On the latter the two blocks of the bimodal compact distribution merge and $\tilde{g}_c(\omega)$ becomes unimodal with one bigger block of width $\tilde\Delta=2\Delta$ such that $K_c=8\Delta/\pi=4\tilde\Delta/\pi$ does not violate the critical value found previously.

Alternatively, one can confirm the critical coupling strength by substituting $\tilde{g}_c$ with the limit of the bimodal rational distribution $\tilde{g}_n$.
As $\lim_{n\to\infty} \tilde{g}_n''(0) = 0$ (as well as all higher derivatives at $\omega=0)$, \eqref{eq:bimodal_selfcons_approx} becomes
\begin{equation}\label{eq:helpbimodal1}
1=  \frac{\pi}{2} K \tilde{g}_c(0) = \frac{K}{2\Delta} n \sin\big(\tfrac{\pi}{2 n}\big) \frac{\Delta^{2 n} }{\Delta^{2n} + \Omega^{2n}} \ 
\end{equation}
for finite $n$.
Solving \eqref{eq:helpbimodal1} for $K$ and taking the limit $n\to\infty$, we find
\begin{equation}\label{eq:bimod_self_cons_Kc}
K_c = \lim_{n\to\infty} \frac{2\Delta}{n \sin\big(\tfrac{\pi}{2 n}\big) }\frac{\Delta^{2n} + \Omega^{2n}}{\Delta^{2 n} } = \begin{cases} 4\Delta/\pi , \ \Delta>\omega_0\\
8\Delta/\pi , \ \Delta=\omega_0\\
\infty, \hspace{.2cm} \text{ otherwise}. \end{cases}
\end{equation}
Neither of the two approaches can detect the Hopf bifurcation boundary when $\tilde{g}_c$ actually is bimodal, that is, below the diagonal $\Delta=\omega_0$.
While the self-consistency argument does not lead to concise results with respect to bifurcation boundaries on the right of the diagonal, we can still rely on the reasoning around (\ref{eq:bimodalHopf}~\&~\ref{eq:couplingscalingfactor}).
According to them the Hopf bifurcation will occur at $K_c = 2\kappa_c \Delta$ with $\kappa_c = \lim_{n\to\infty} \kappa_n = 4/\pi$ for $\Delta < \omega_0$.

\subsubsection{Partially synchronized solutions}
Away from the incoherent solution $R=0$, one can solve \eqref{eq:self-consK} by inserting the definition of the bimodal compact distribution $\tilde{g}_c(\omega)=\tilde{g}^{(\Delta,\omega_0)}(\omega)$.
In particular, $\tilde{g}_c(\omega)$ does not vanish when $|KR\sin \theta| \in [\omega_0-\Delta,\omega_0+\Delta|$.
Consequently, and after some straightforward algebra, we find
\begin{equation}
\begin{aligned}
\frac{4\Delta}{K} &= \arcsin\big( \tfrac{\omega_0+\Delta}{KR}\big)+ \arcsin\big( \tfrac{\Delta-\omega_0}{KR}\big)\\
&\hspace{-.4cm} + \tfrac{\omega_0+\Delta}{KR}\sqrt{1-\big(\tfrac{\omega_0+\Delta}{KR}\big)^2}  + \tfrac{\Delta-\omega_0}{KR}\sqrt{1-\big(\tfrac{\Delta-\omega_0}{KR}\big)^2} \ .
\end{aligned}
\label{eq:bimod_self_consfp}
\end{equation}
As all parameters are real-valued, in particular the coupling strength $K$, the self-consistency equation \eqref{eq:bimod_self_consfp} is only valid if
$$
KR \geq \omega_0+\Delta  \text{ and } KR \geq \Delta-\omega_0 \ \Rightarrow \ KR \geq \omega_0+\Delta \ .
$$
We thus find critical values for $K_c$ and $R_c$ by inserting $K_cR_c=\omega_0+\Delta$ into \eqref{eq:bimod_self_consfp}:
\begin{equation}\label{eq:bimod_self_cons_Kconditions}
\frac{4\Delta}{K_c} = \tfrac{\pi}{2} + \arcsin\big( \tfrac{\Delta-\omega_0}{\Delta+\omega_0} \big) + \tfrac{\Delta-\omega_0}{\Delta+\omega_0}\sqrt{1-\big( \tfrac{\Delta-\omega_0}{\Delta+\omega_0} \big)^2 }.
\end{equation}
For $\omega_0\to 0$, which is the unimodal compact distribution of width $\tilde{\Delta}=\Delta$, this results in $R_c = \pi/4$ and $K_c = 4\Delta/\pi$.
For $\omega_0 \to \Delta$ we have again a unimodal compact distribution, now of width $\hat{\Delta}=2\Delta$.
There we retain the critical values $R_c = \pi/4$ and $K_c = 8\Delta/\pi = 4\hat{\Delta}/\pi$.
Note that for $\omega_0\leq \Delta$, the critical coupling $K_c$ smoothly increases from $4\Delta/\pi$ to $8\Delta/\pi$.
By contrast, the critical coupling denoting the stability boundary of the incoherent state remains constant for all $\omega_0\leq \Delta$.
This suggests that either no or multiple fixed point solutions exist between the two stability boundaries for the incoherent and the partially synchronized state.

\subsection{Complete bifurcation diagram following numerical analysis}
\begin{figure}[htb]
\centering
\setbox1=\hbox{\includegraphics[width=.75\columnwidth]{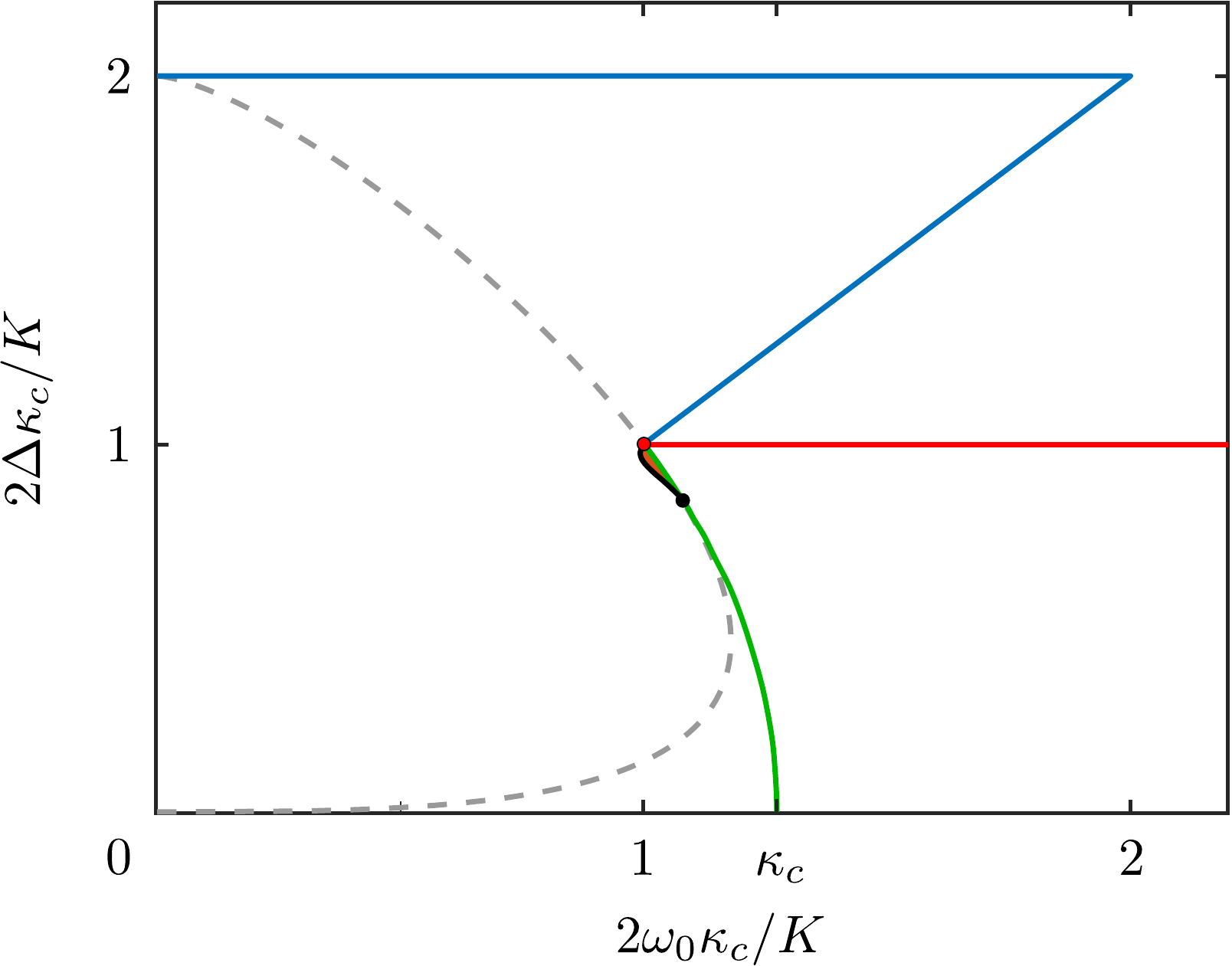}}
 \includegraphics[width=.75\columnwidth]{compactBifurcationDiagram1}\llap{\makebox[.825\wd1][l]{\raisebox{1.1cm}{\frame{\includegraphics[height=2cm]{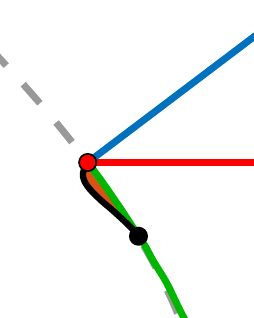}}}}}
\caption{Bifurcation diagram for the bimodal compact distribution.
The inset shows a zoom into the bistability area with $\tilde{\omega} \in [0.86,1.26]$ and $\tilde{\Delta} \in [0.68,1.34]$. Color coding as in Fig.~\ref{fig:martensAnalysis}.
The gray dashed line is the stability boundary of the partially synchronized fixed point solution according to \eqref{eq:bimod_self_cons_Kconditions}.}
\label{fig:compactBifDiagram}
\end{figure}

To provide a comprehensive picture of the Kuramoto model with a bimodal compact frequency distribution and its different dynamical regimes beyond mere fixed point solutions, we employed numerics.
We ran detailed numerical simulations of $N=100,000$ oscillators for $T=500$ seconds employing a Runge-Kutta45-ODE solver with adaptive step size in MATLAB \cite{MATLAB:2017}.
Scanning the entire $\Delta - \omega_0$ parameter space, we started the simulations from different (macroscopic) initial conditions in order to test for bistability.
The resulting bifurcation diagram is shown in Fig.~\ref{fig:compactBifDiagram}.
Using the same color coding as before, the similarity to the bifurcation diagrams for the bimodal rational distributions is striking.

\subsubsection{Above the diagonal $\Delta=\omega_0$}
For parameter values $\Delta \geq \omega_0$, the compact `bimodal' frequency distribution $g_c(\omega)$ is effectively unimodal, see also Fig.~\ref{fig:bimodregions}.
The results of the self-consistency approach in the previous \textit{Sub-section}~\ref{subsec:self-cons_bimod}, in particular \eqref{eq:bimod_self_cons_Kc} and \eqref{eq:bimod_self_cons_Kconditions}, equally hold above and below the diagonal $\Delta=\omega_0$.

According to \eqref{eq:bimod_self_cons_Kc} and confirmed by our numerical simulations, the transcritial bifurcation boundary of the incoherent solution is a piecewise linear curve (blue) with a sharp edge at the diagonal $\Delta=\omega_0$.
The gray dashed line depicts the stability boundary of the partially synchronized, phase locked solution predicted by \eqref{eq:bimod_self_cons_Kconditions}.
As to time-asymptotic solutions, the gray curve does not present the boundary for any particular dynamical regime.
Yet, when considering the relaxation dynamics towards the (unique) stable fixed point from different initial conditions, there is a slight qualitative difference visible in the order parameter dynamics.
Within the gray curve, all (macroscopic) initial conditions show the same relaxation behavior towards the unique fixed point.
Beyond that curve but within the dynamical regime of only one fixed point, the relaxation dynamics depend on the initial conditions.
Fig.~\ref{fig:dynamicsTop} shows the exemplary evolution of the global (black) and local (red/blue) real-valued order parameters for parameter values slightly to the right of the gray curve ($\Delta=1.6, \omega_0=0.7$).
When starting from initial conditions close to incoherence (left), the order parameter dynamics converges exponentially to the fixed point solution, which suggests the fixed point to be a stable node.
On the other hand, for initial conditions close to full synchrony, we see a clear ringing effect, that is, the dynamics feature damped oscillations around the (same) fixed point, giving rise to the assumption that it is a stable focus.
As such, the gray curve does indeed provide another stability boundary.
This additional complexity of the collective behavior seems to be inherent to the infinite-dimensional dynamics that governs the evolution of the order parameter(s); see also \textit{Sub-section}~\ref{subsec:compact_bimod_remarks}.

\subsubsection{Below the diagonal $\Delta=\omega_0$}
The Hopf bifurcation boundary (red curve in Fig.~\ref{fig:compactBifDiagram}) at $K_c=2\kappa_c\Delta$ according to \eqref{eq:bimodalHopf} is also confirmed numerically, which adds to the resemblance of the overall bifurcation structure to that for the flat bimodal but smooth (rational) distributions $\tilde{g}_n(\omega)$ with $n<\infty$.
A major difference, however, is the disappearance of the yellow bistability region above the Hopf curve, cf. Figs.~\ref{fig:martensAnalysis} and \ref{fig:bifdiagramsN234}.
That is, there is no longer a coexistence possible between two stable fixed-points on the right of the diagonal $\Delta=\omega_0$.
At the same time this restricts the possible routes to synchronization as the incoherent solutions always has to undergo a Hopf bifurcation first!

\begin{figure}[t]
\centering{
\subfigure[ $\Delta=1.6, \omega_0=0.7$]{\includegraphics[width=.8\columnwidth]{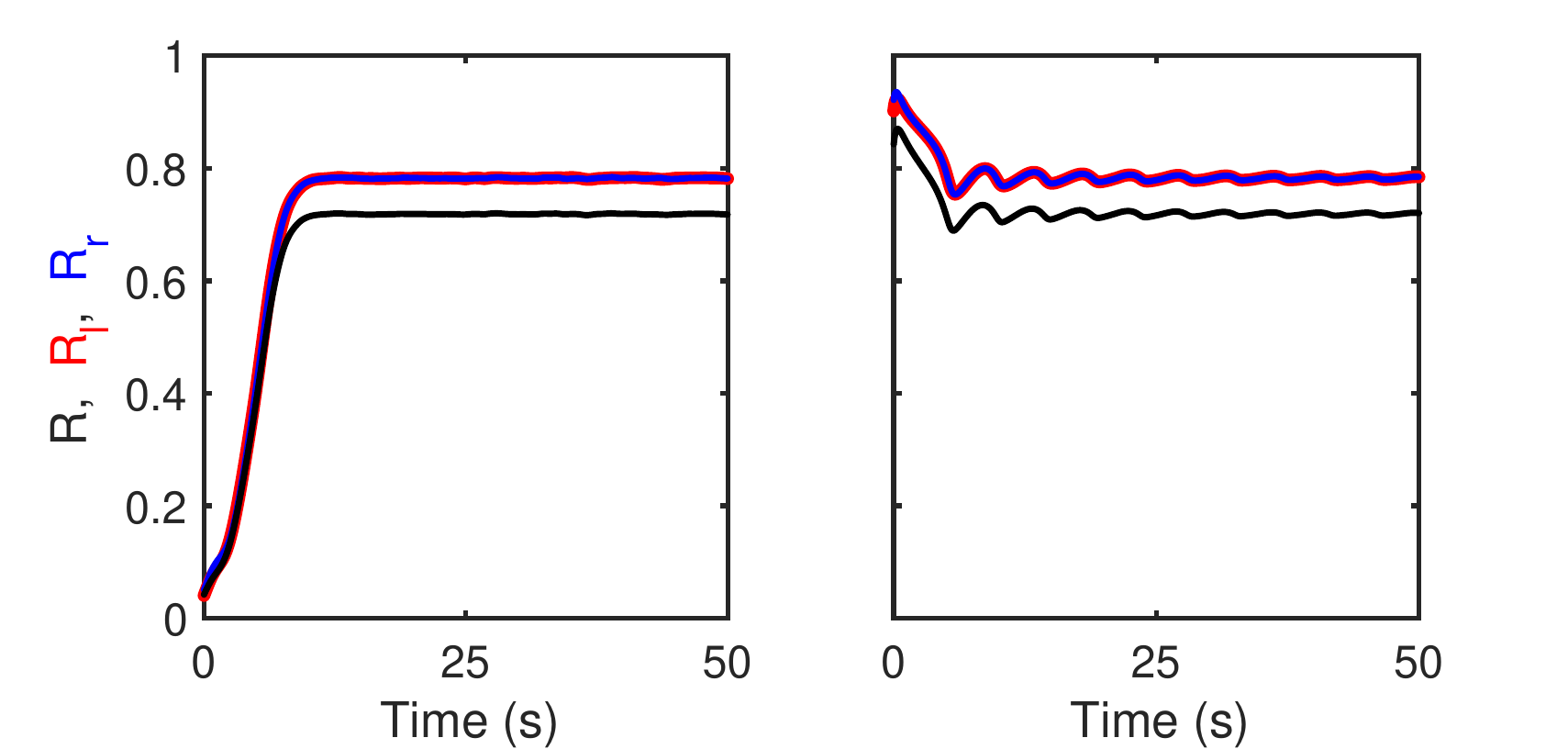}\label{fig:dynamicsTop}}
        
\subfigure[ $\Delta=0.4, \omega_0=1.19$]{\includegraphics[width=.8\columnwidth]{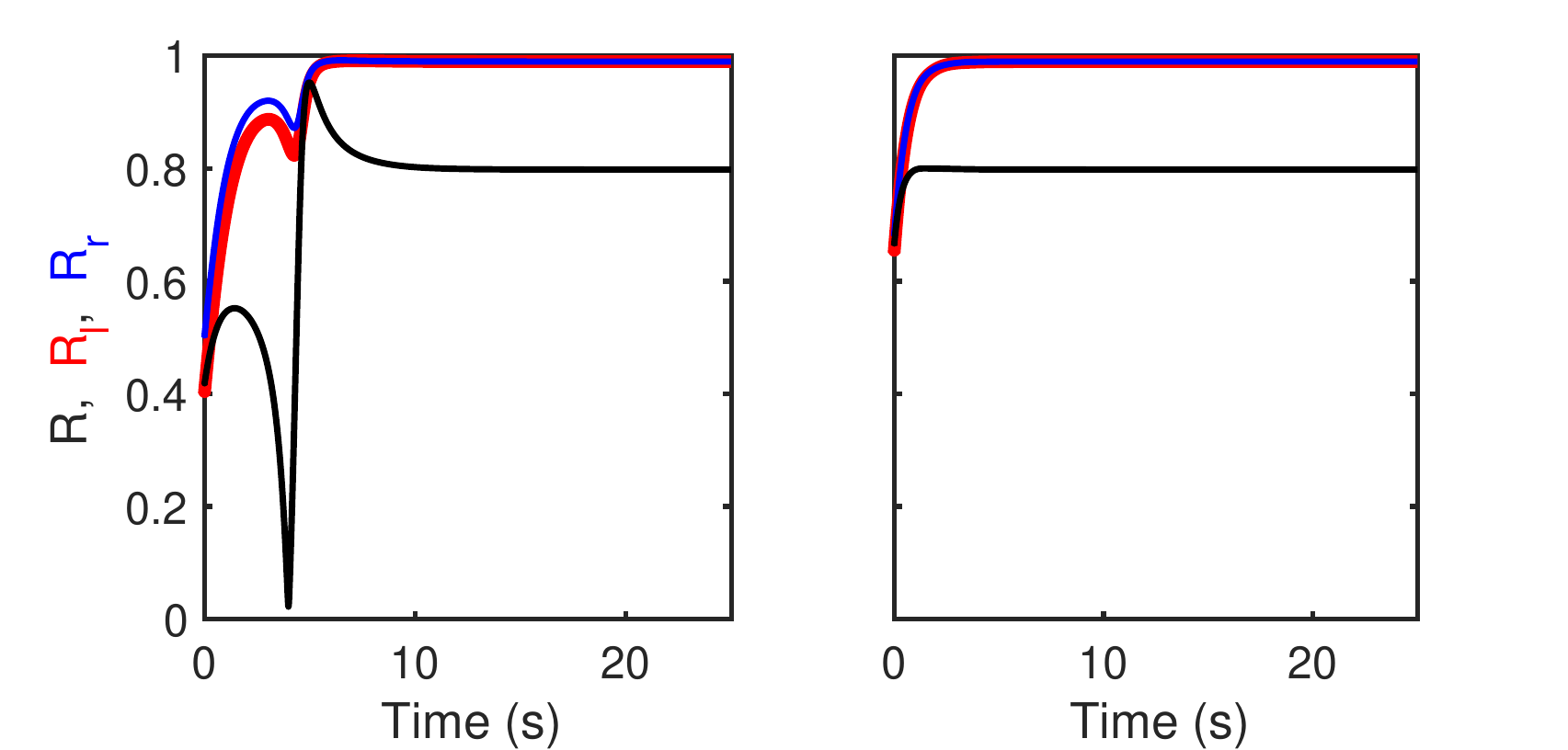} \label{fig:dynamicsLow}}
       
}
\caption{Relaxation dynamics towards the same fixed point from different initial conditions.
Shown are the global (black) and local (red/blue) real-valued order parameters, $R=|z|$ and $R_{l,r}=|z_{l,r}|$, for a network of $N=100'000$ oscillators whose natural frequencies follow a bimodal compact distribution $\tilde{g}^{(\Delta,\omega_0)}_c(\omega)$ with parameter values slightly to the right of the gray stability curve above (a) and below (b) the diagonal $\Delta=\omega_0)$ and $K=2\kappa_c$.}
\label{fig:samesamebutdifferent}
\end{figure}

Beyond the Hopf curve, the bistability region with one attractive oscillating solution coexisting with a stable fixed point solution (brown area) survives in the limit $n\to\infty$.
Remarkably, the saddle node curve (green) coincides in this area with the gray dashed curve following \eqref{eq:bimod_self_cons_Kconditions}.
In particular, the convergence holds between the two co-dimension 2 bifurcation points, the Bogdanov-Takens (red dot) and the saddle-node-loop point (black).
Beyond the saddle-node-loop bifurcation the green curve depicts a SNIC bifurcation.
Moreover, away from the saddle node bifurcation, the gray dashed line describes the same qualitative change of collective behavior as seen in the parameter region above the diagonal, see the exemplary Fig.~\ref{fig:dynamicsLow}.

\subsubsection{A note on the system's $O(2)$ symmetry}
\label{subsec:compact_bimod_remarks}
The last missing piece of the compact bimodal puzzle is to explain the dynamics shown in Fig.~\ref{fig:samesamebutdifferent} and the dashed curve \eqref{eq:bimod_self_cons_Kconditions} found along the self-consistency approach.
Capitalizing on the inherent $O(2)$ symmetry of the Kuramoto model with a symmetric frequency distribution, Crawford identified the simultaneous emergence of both `standing waves' and `traveling waves' \cite{Crawford1994}.
While standing waves depict macroscopic oscillations with a time varying amplitude $R(t)$ of the Kuramoto order parameter $z=R\mathrm{e}^{i\Psi}$, traveling waves are defined as stationary solutions with a central frequency $\Omega \neq 0$ different from the central frequency of the co-rotating frame (where $\Omega=0$) \cite{Iatsenko2013,Petkoski2013meanfield}.
Such traveling wave states can become stable for asymmetric frequency distributions \cite{Petkoski2013meanfield}, but typically they are unstable \cite{Crawford1994}.
They persist for larger coupling strengths than stable standing waves \cite{bonillaperezvicente1998}.
This agrees with our findings below the diagonal $\Delta=\omega_0$, where the gray dashed curve is left of the green SNIC curve, that is, at larger coupling strengths.
Previous results on the bimodal Lorentzian in both sum and difference formulations also confirm the existence of such unstable traveling wave structures, see, e.g., the gray dashed curve in Fig.~\ref{fig:martensAnalysis}, although these traveling waves are more subtle as they only exist outside the reduced planes of `relevant', i.e. attractive, dynamics \cite{MartensExactResults2009,PazoMontbrio2009existence}.

When considering parameter values at the unimodal-bimodal boundary and above the diagonal $\Delta=\omega_0$, explanations for the gray dashed curve become less rigorous.
In the Lorentzian cases, the gray dashed curves coincide with the transcritical (see Fig.~\ref{fig:martensAnalysis}) and pitchfork bifurcations (see Fig.~4 in \cite{PazoMontbrio2009existence}), respectively.
For the compact bimodal distribution, however, there is a prominent gap between the blue transcritical and the gray dashed curves, see Fig.~\ref{fig:compactBifDiagram}.
Together with the dynamics depicted in Fig.~\ref{fig:samesamebutdifferent}, this strongly suggests the existence of unstable traveling wave solutions.
Following Crawford, there may emerge traveling waves also at the steady-state bifurcation (which is the transcritical bifurcation in our notation), especially when breaking the reflection-symmetry of the system, see Section~3.1.5 on ``Perturbing $O(2) \to SO(2)$'' in \cite{Crawford1994}.
Obviously, the compact bimodal distribution $\tilde{g}_c(\omega)$ allows for reflections $(\theta,\omega) \mapsto (-\theta,-\omega)$ about the origin.
Does a break of (reflection-)symmetry occur at the sharp edges of the step-like frequency distribution when $\Delta > \omega_0$ similar to the effects of time delay considered by Montbri\'o and co-workers in \cite{Montbrio2006}?
We already know that the Kuramoto model with compact frequency distributions has some special dynamical properties, e.g., the sub-exponential relaxation dynamics to the incoherent solution \cite{StrogatzMirolloMatthews1992,Strogatz2000} or the scaling behavior in the critical regime \cite{Pazo2005,Skardal2018}.
While this may hint at the peculiarities of the infinite-dimensional order parameter dynamics between the gray dashed curve and the transcritical bifurcation, a rigorous mathematical proof is highly desirable.


\section{Explosive or continuous synchronization?}
\label{sec:explosiveOrcontinuous}
With the loss of the (yellow) bistability region in Fig.~\ref{fig:compactBifDiagram}, the collective dynamics of the Kuramoto model with a bimodal compact frequency distribution appears to be less complex than those systems with bimodal rational frequency distributions.
However, as illustrated in Fig.~\ref{fig:samesamebutdifferent}, the actual collective behavior of the bimodal compact network is not as simple as it seems.

As a final point we investigate the nature of transitions from incoherence towards synchronous collective behavior in the different bimodal networks.
We would like to recall that the sharper the edges of the unimodal rational frequency distribution become, that is, for increasing $n>1$, the more exposed is the discontinuous character of the phase transition.
A natural question is which role these \emph{first-order phase transitions} play in networks with flat bimodal frequency distributions.
While the bimodal Lorentzian network exhibits first-order phase transitions \cite{MartensExactResults2009}, one may ask whether finite plateaus in the frequency distribution have a catalyzing, or rather a counteracting, effect on the discontinuous, \emph{explosive synchronization} properties of the network.

\begin{figure}[b!]
\centerline{\includegraphics[width=.85\columnwidth]{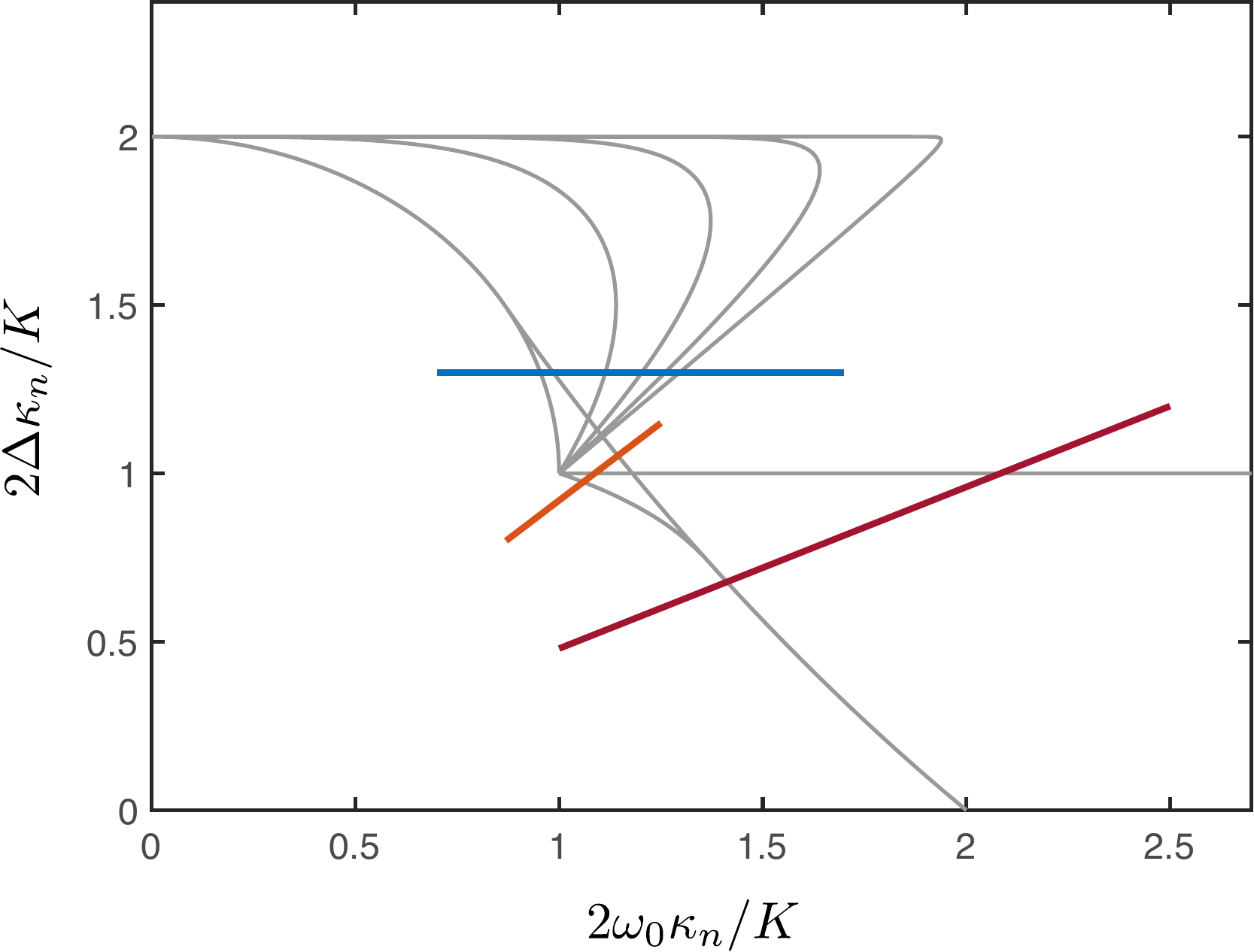}}
\caption{Paths of varying coupling strength. The blue line crosses the transcritical bifurcation boundary. Here we fix $\Delta=1.3$ and $K=2 \kappa_n$ while varying $\omega_0 \in [0.7,1.7]$. The red line crosses the Hopf bifurcation when varying $K \in 2\kappa_n [1,4]$ for $\Delta=1.25, \omega_0=2.5$. The orange curve passes through the bistable region when varying $K\in 2\kappa_n[1,3/2]$ and fixing $\Delta=1.15, \omega_0=1.25$.}
\label{fig:varyCouplingOverview}
\end{figure}

\begin{figure}[!bh]
\centering{
\includegraphics[width=.98\columnwidth]{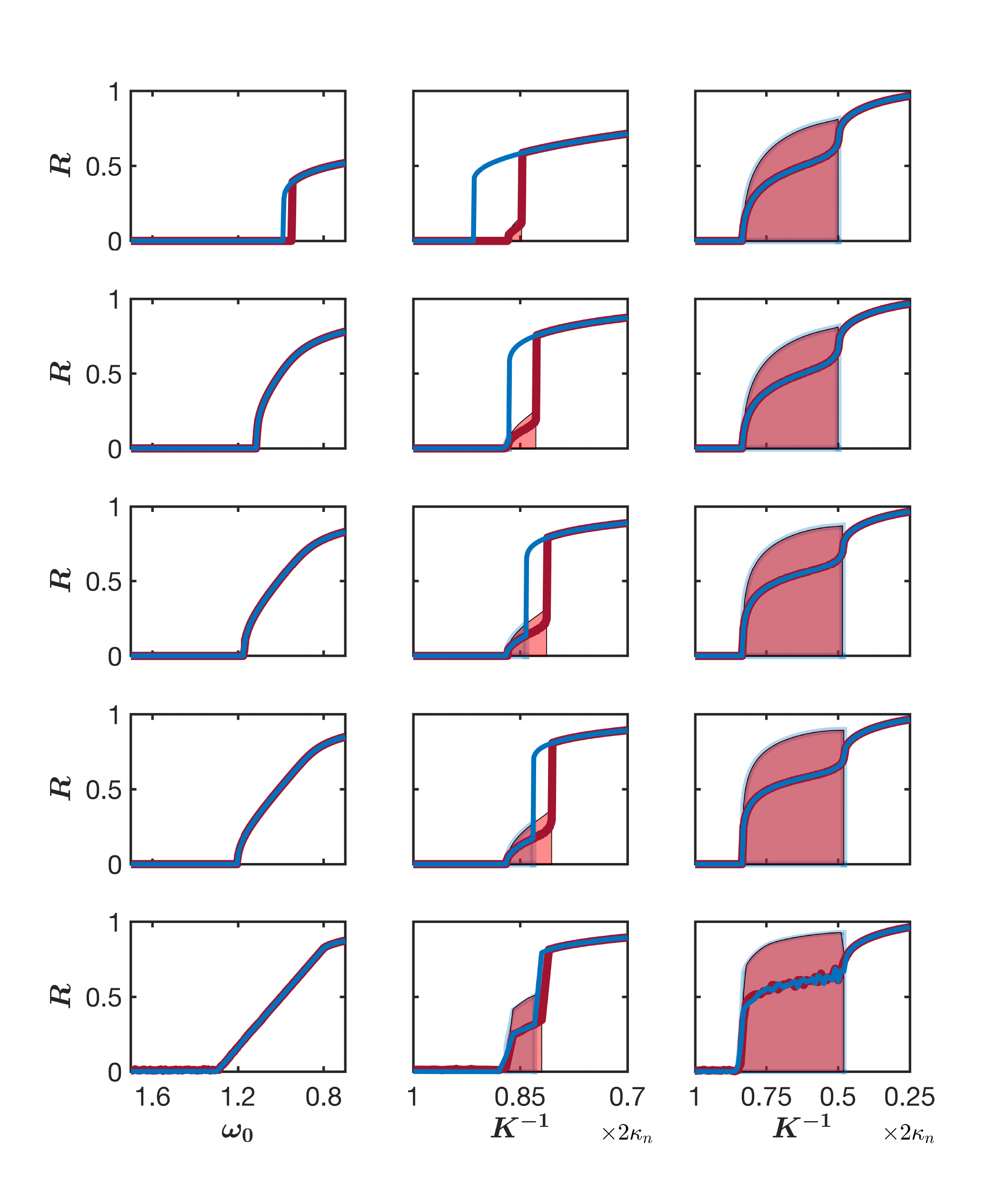}
}
\caption{Transitions to synchrony in networks with bimodal frequency distributions $g_n(\omega)$ with $n=1,2,3,4,c=\infty$ from top to bottom.
The rows represent the change of collective behavior when varying the respective bifurcation parameters along the colored curves in Fig.~\ref{fig:varyCouplingOverview}.
Thick curves denote the mean degree of phase synchronization $R$ and shaded regions denote maximal amplitudes of collective oscillations.
Colors (Blue/red) represent the behavior for an adiabatic change (increase/decrease) of the bifurcation parameter.
Left row: crossing the transcritical bifurcation, which changes from a first-order to a second-order phase transition for increasing $n$.
Middle: crossing the bistability area with decreasing size of the corresponding hysteresis cycle.
Right: crossing the Hopf bifurcation, which changes from a second-order to a first-order phase transition.
}
\label{fig:syncTransitions}
\end{figure}

We studied the exemplary routes to synchronization.
In particular, we varied the parameters along three colored lines depicted in Fig.~\ref{fig:varyCouplingOverview}:
The blue line corresponds to crossing the transcritical bifurcation boundary at the edge of unimodality, the red line denotes the transition through the Hopf bifurcation, and the orange line passes through the saddle-node bifurcation into a region of bistability.
The latter two lines can be parametrized by the coupling strength $K$ for fixed parameters $\Delta$ and $\omega_0$, whereas we fixed $K$ and $\Delta$ and varied the distance between the bimodal peaks $\omega_0$ along the first curve.
In Fig.~\ref{fig:syncTransitions}, we display the different transition behavior along the colored lines and plot the absolute value of the global Kuramoto order parameter $R = |z| = \tfrac{1}{2} |z_l + z_r| $ as a function of the respective bifurcation parameter ($\omega_0$ along the blue line, and the coupling strength $K$ along the red and orange lines in Fig.~\ref{fig:varyCouplingOverview}).
To identify possible hysteresis cycles, we varied the bifurcation parameters adiabatically.
The blue curves represent the order parameter for increasing bifurcation parameter values, whereas we decreased the bifurcation parameters for the red curves.
The error bars denote minimum and maximum amplitudes of oscillatory order parameter dynamics.

The left row of Fig.~\ref{fig:syncTransitions} displays the synchronization behavior when crossing the transcritical bifurcation (blue line), the middle row that when passing through the bistability region, and the right row when crossing the Hopf curve.
Along the columns we increased the parameter $n=1,2,3,4, \infty$.

When analyzing the bifurcation diagrams, the parameter region of bistability shrinks for increasing $n$.
Bistability is characterized by the hysteresis cycle in the dynamics of $R$, which is consistently displayed in the middle row of Fig.~\ref{fig:syncTransitions}.
The hysteresis cycles shift to the right for larger $n$, which indicates the increasing slope of the saddle-node curves in the bifurcation diagrams in Fig.~\ref{fig:bifdiagramsN234}.
The cycles become narrower from top to bottom which corresponds to the shrinking bistability area.

As to the transcritical bifurcation (left row), the offset of synchronization moves to the left for larger $n$, which is in accordance with the transcritical curves in the bifurcation diagrams in Fig.~\ref{fig:stabilityboundaries} that move closer to the diagonal $\Delta=\omega_0$.
In contrast to the unimodal frequency distributions, the phase transition for the bimodal Lorentzian frequency distribution, $n=1$, is discontinuous, whereas for larger $n$ the phase transition is clearly of second-order, i.e. continuous.
In case of the bimodal compact distribution, the transcritical character of the bifurcation becomes evident, as the order parameter $R$ increases linearly from zero.

When crossing the Hopf bifurcation, the nature of phase transitions changes in the exactly opposite direction as in the transcritical case, see the right row of Fig.~\ref{fig:syncTransitions}.
The bimodal Lorentzian network features the emergence of small amplitude oscillations (as denoted by the error bars) close to the critical coupling strength.
For increasing $n$, however, the amplitudes of oscillation grow rather abruptly and resemble the discontinuous transitions known from the unimodal networks.
Finally, for the bimodal compact frequency distribution, the first-order phase transition becomes apparent and the amplitudes in the region of collective oscillations are largest.
This abrupt onset of global oscillations with a finite amplitude also explains why the self-consistency approach as outlined in the previous sub-section does not succeed in predicting the Hopf bifurcation for the bimodal compact network.

Apparently, there is indeed a change of synchronization behavior for the Kuramoto model with flat frequency distributions.
While the hysteresis cycle (between a partially synchronized solution and low-amplitude oscillations) including first-order phase transitions is a dominant feature for all bimodal frequency distributions, behavior switches when crossing the transcritical and the Hopf bifurcation boundaries.
At the onset of global oscillations the (mean) order parameter jumps discontinuously to a higher finite value for larger $n>1$.
By contrast, the phase transition via the transcritical bifurcation curve becomes smoother and converges to the typical straight line characteristic for transcritical bifurcations.


\section{Conclusion \& Discussion}
\label{sec:conclusion}
The primary aim of this study was to investigate the collective dynamics of the Kuramoto model with flat bimodal frequency distributions.
While both bimodal (smooth) frequency distributions \cite{bonilla1992nonlinear,MartensExactResults2009} as well as flat unimodal distributions \cite{Pazo2005} lead to explosive synchronization phenomena, the effects on the network dynamics when combining bimodal and flat frequency distributions were largely unclear.
Thanks to the recent introduction of rational frequency distributions $g_n(\omega)$, which approximate the compact (uniform) distribution in the limit $n\to \infty$ \cite{Skardal2018}, we could derive the governing low-dimensional dynamics of the network's order parameter by employing the OA ansatz \cite{OttAntonsen2008}.
We were able to extrapolate the analytic insights for finite $n=1,2,\dots$ to the Kuramoto model with a bimodal compact frequency distribution.
Using also Kuramoto's original self-consistency argument \cite{Kuramoto1984}, we determined the backbone of the bifurcation diagram for the compact bimodal network, see Fig.~\ref{fig:compactBifDiagram}.
Numerical simulations helped to fill the remaining gaps in the bifurcation structure.

The main result of our analysis is the overall similarity of the bifurcation diagrams for all bimodal flat frequency distributions.
Yet, a decisive feature is that even close to the unimodal-bimodal border the transition from incoherence to synchrony always appears through collective oscillations.
The route to synchronization is restricted via this oscillatory state, which is in contrast to smooth bimodal frequency distributions -- at least when they are compounded as the sum of two unimodal distributions with infinite support each (as is the case considered by Martens and co-workers \cite{MartensExactResults2009}).
Still, the compact bimodal distribution represents a natural limit between the sum and difference formulations of bimodal distributions.
The difference of two Lorentzians has been considered by Paz\'o and Montbri\'o, who found a similar disappearance of the intermediate step of bistability on the route to synchronization \cite{PazoMontbrio2009existence}.
We can conclude that the compact bimodal case thus defines a natural link from one to another route to synchronization.

Another important feature of the compact bimodal Kuramoto model is the occurrence of first-order phase transitions at the onset of global oscillations, that is, at the Hopf bifurcation.
A small but non-negligible area of bistability is inherent to all networks, where abrupt transitions from one to another dynamic behavior define the edges of the corresponding hysteresis cycle.
At these discontinuities as well as at the Hopf bifurcation boundary it may be interesting to investigate in future studies the scaling properties of the order parameter.
Skardal already offered some rigorous results for the order parameter behavior close to the onset of synchronization for unimodal rational frequency distributions \cite{Skardal2018}.
While his results apply only to the continuum limit of infinitely many oscillators, it is an open problem how finite-size effects shape the scaling properties of the order-parameter in the critical regime for the compact bimodal network; see \cite{Pazo2005,coletta2017} for results on the compact unimodal frequency distribution, and, e.g., \cite{buice2007,ghosh2013relaxation,peter2017} for more general results on finite-size Kuramoto models with compact frequency distributions.

An intriguing line of research is to identify the role of explosive synchronization on chimera states \cite{panaggio2015chimera}.
Previous work established the direct link between the Kuramoto model with a symmetric bimodal frequency distribution and two coupled Kuramoto models each with a unimodal frequency distribution \cite{PiDeDa2016}.
This setup resembles the first analytic account on chimera states by Abrams and co-workers \cite{abrams2008solvable}, which has recently been generalized to coupled Kuramoto networks with distributed frequencies \cite{kotwal2017connecting}.
Numerical results already hint at chimera-like behavior when allowing for compact frequency distributions in this two-population setup \cite{zhang2016bridging}.
Our results will certainly aid explaining the underlying mechanisms for this peculiar network behavior and shed light on the connection between chimera states and explosive synchronization.

\begin{acknowledgements}
\noindent
This project has received funding form the European Union's Horizon 2020 research and innovation program under the Marie Sk{\l}odowska-Curie grant agreement \#642563 (COSMOS).
\end{acknowledgements}




 \newcommand{\noop}[1]{}

\end{document}